\documentclass[final,onefignum,onetabnum]{SISC/siamart190516}
\usepackage[utf8]{inputenc}
\usepackage{graphicx}
\usepackage{amsmath}
\usepackage{amssymb}
\usepackage{subfig}
\usepackage{enumitem,kantlipsum}
\usepackage[algo2e,boxed,ruled]{algorithm2e}
\usepackage{tcolorbox}
\usepackage{hyperref}

\graphicspath{{./images}}

\newcommand{\bfx}{\mathbf{x}}

\newcommand{\bfz}{\mathbf{z}}

\newcommand{\bfd}{\mathbf{d}}

\newcommand{\bfq}{\mathbf{q}}

\newcommand{\bfm}{\mathbf{m}}
\newcommand{\bfu}{\mathbf{u}}

\newcommand{\bfJ}{\mathbf{J}}
\newcommand{\bfI}{\mathbf{I}}
\newcommand{\bfP}{\mathbf{P}}
\newcommand{\bfZ}{\mathbf{Z}}
\newcommand{\bfA}{\mathbf{A}}
\newcommand{\bfX}{\mathbf{X}}
\newcommand{\bfW}{\mathbf{W}}
\newcommand{\bfD}{\mathbf{D}}
\newcommand{\bfQ}{\mathbf{Q}}
\newcommand{\bfR}{\mathbf{R}}
\newcommand{\bfT}{\mathbf{T}}
\newcommand{\bfH}{\mathbf{H}}

\newcommand{\helm}{\mathcal {H}}

\newcommand{\vecsf}{{\sf {vec}}}
\newcommand{\reduced}{{\sf {_{Reduced}}}}
\newcommand{\extsrc}{{\sf {_{ExtSrc}}}}
\newcommand{\lowrankextsrc}{{\sf {_{LowRankExtSrc}}}}
\newcommand{\lowrankextsimsrc}{{\sf {_{LowRankExtSimSrc}}}}
\newcommand{\reducedsplit}{{\sf {_{ReducedSplit}}}}
\newcommand{\constrained}{{\sf {_{Constrained}}}}
\newcommand{\penalized}{{\sf {_{Penalized}}}}

\newcommand{\true}{{\sf {_{true}}}}
\newcommand{\obs}{{\sf {_{obs}}}}

\title{Full waveform inversion using extended and simultaneous sources\thanks{Corresponding author: Eran Treister. \funding{This research was partially supported by The Israel Science Foundation (grant No. 1589/19).}}}
\author{Sagi Buchatsky\footnotemark[2] and Eran Treister\thanks{Department of Computer Science, Ben-Gurion University of the Negev, Beer Sheva, Israel. ({\tt erant@cs.bgu.ac.il, sagibu@post.bgu.ac.il})}}
\date{}

\begin{document}

\maketitle

% REQUIRED
\begin{abstract}
PDE-constrained optimization problems are often treated using the reduced formulation where the PDE constraints are eliminated.
This approach is known to be more  computationally feasible than other alternatives at large scales. However, the elimination of the constraints forces
the optimization process to fulfill the constraints at all times. In some problems this may lead to a highly non-linear objective, which is hard to solve.
An example to such a problem, which we focus on in this work, is Full Waveform Inversion (FWI), which appears in seismic exploration of oil and gas reservoirs, and medical imaging. In an attempt to relieve the non-linearity of FWI, several approaches suggested to expand the optimization search space and relax the PDE constraints. This comes, however, with severe memory and computational costs, which we aim to reduce. In this work we adopt the expanded search space approach, and suggest a new formulation of FWI using extended source functions. To make the source-extended problem more feasible in memory and computations, we couple the source extensions in the form of a low-rank matrix.
This way, we have a large-but-manageable additional parameter space, which has a rather low memory footprint, and is much more suitable for solving large scale instances of the problem than the full rank additional space. In addition, we show how our source-extended approach is
applied together with the popular simultaneous sources technique---a stochastic optimization technique that significantly reduces the computations needed for FWI inversions. We demonstrate our approaches for solving FWI problems using 2D and 3D models with high frequency data only.
\end{abstract}


\pagestyle{myheadings} \thispagestyle{plain} \markboth{Sagi Buchatsky and Eran Treister}{FWI USING EXTENDED AND SIMULTANEOUS SOURCES}

\numberwithin{equation}{section}
\section{Introduction}
Many computational science applications involve parameter estimation of partial differential equations (PDEs), known as inverse problems \cite{haber2014computational,vogel2002computational,somersalo2004statistical}. One challenging inverse problem is \emph{Full Waveform Inversion} (FWI), where we aim to determine the wave velocity, density, and possibly other parameters of a heterogeneous medium, given sources and waveform observations at receiver locations on its boundary \cite{metivier2017review}. Inverse problems of this type arise in seismic exploration of oil and gas reservoirs, earth sub-surface mapping, ultrasound imaging \cite{bernard2017ultrasonic}, optical diffraction tomography \cite{soubies2017efficient}, brain imaging \cite{guasch2020full} and more.

 The estimation of the velocity model is usually performed by fitting numerically simulated data to observed field data. This results in an optimization problem that needs to be solved by an iterative descent algorithm,
which gradually reduces the misfit between the simulated and field data. This data fitting problem is usually accompanied by a suitable regularization that aims to introduce prior information on the estimated coefficients \cite{taran}.  In this paper we focus on FWI in the context of seismic exploration, where the wave velocity of the earth subsurface is estimated. This problem has gained popularity in the last decade with the advances in data acquisition techniques, computing power and numerical algorithms
\cite{pratt1999,EpanomeritakisAkcelikGhattasBielak2008,krebs09ffw,shin2009waveform,biondi2014simultaneous,van20143d,van2013mitigating,metivier2017review}.

Solving the FWI problem is challenging in two main aspects, both of which may be relevant to other inverse problems. First, it is an ill-posed and highly non-convex problem, especially if the sources and receivers are placed on the same surface. Then the problem typically has
multiple minima, and convergence to a local minimum leads to a non-plausible estimated model. Many approaches have been proposed for solving FWI, however, it is still considered difficult to solve, as solutions techniques can be unstable, converging to local minima in many scenarios. Some recent works suggest to complement FWI with other more stable inverse problems such as travel time tomography \cite{liu2017joint,JointEikFWI17} or electromagnetic inversion \cite{haber2016obtaining}. Other approaches can be found in \cite{biondi2014simultaneous,Li2016extrapolation}. Such techniques are useful in some cases, but still, FWI is considered challenging and not robust enough in real life.

In addition to being ill-posed, FWI is highly computationally expensive, both in memory usage and computations. The problem typically involves a large number of sources and several frequencies, and requires multiple solutions of the forward problem to simulate the data, each of which is challenging by itself. These numerical simulations are either obtained by propagating the wave equation (in time domain), or by solving the Helmholtz equation (in frequency domain). Both options introduce computational challenges due to the scale and properties of the problem. The aim of this work is to help dealing with both of the aforementioned challenges---non-linearity and computational cost---and make the FWI solution process more robust, but also more computationally feasible at the same time, involving less forward simulations.

\emph{Relieving the non-linearity by enlarging the search-space} Our first objective in this work is to relieve some of non-linearity in FWI. Recent works \cite{aghamiry2018improving,huang2018volume,van2016} have demonstrated that relaxing the PDE constraints and enlarging the search-space may allow us to avoid local minima, and find plausible solution of FWI from rather arbitrary initial guesses. This can be obtained by introducing the full waveforms as variables, or extending the point-wise sources to the full domain.  However, this comes with a severe memory footprint in 3D, coming from the introduction of the full waveforms or extended sources as variables of the optimization problem. Both options introduce several hundreds (or more) vectors of unknowns to keep track of and iterate on as part of the solution process, each at the size of the 3D domain.

In our first contribution in this paper we suggest a new formulation of FWI that has the additional search parameters in the form of \emph{low-rank and sparse} extended source functions. This way, we relax the PDE constraints, while at the same time we keep the inversion manageable in memory (comparable to the reduced FWI version), by using low-rank source extensions and sparsity promoting penalties.

\emph{Reducing computational complexity}
To further reduce the computational complexity of the extended inversion problem, we adopt a stochastic optimization technique called \emph{simultaneous sources} \cite{haber2012effective}.
In this approach we reduce the number of sources involved in each iteration of the optimization, by applying a randomized trace estimation technique \cite{avron2011randomized} to the data misfit term. While standard stochastic techniques are based on random subsets of the misfit terms (sources), in the simultaneous technique we project all the sources onto a smaller dimension using a random matrix. This way, the sources are randomly mixed, and the inversion is not biased towards any geometry of the sources' locations as in the subset-based methods \cite{haber2012effective}.
Therefore the simultaneous sources technique was found to be more effective than the subset-of-sources technique in general PDE-constrained optimization like DC-resistivity  \cite{roosta2014stochastic,roosta2015improved}, and FWI in particular  \cite{van2011seismic}.

However, as it is, the simultaneous sources technique is not suitable for the experimental setting of typical real-life FWI scenarios. That includes common settings where, for example, each source only affects part of the receivers, or sporadically missing data due to malfunctioning receivers, or cases where the measurement noise parameters (the co-variance matrices) are not similar for all the sources. The particular case of sporadically missing data was addressed in \cite{roosta2014data} using a smooth data completion regularization for the DC-resistivity problem. This solution, however, is not suitable for FWI, for which the data is not smooth. Recently, the work \cite{Liu_2018} adjusted the simultaneous sources technique in the case of missing data by splitting the data misfit function into two terms, so that one is suitable to the simultaneous sources method and the other is easy to handle computationally (we elaborate on this later). This was accompanied by a low-rank regularization to the data term, which is suitable for FWI, in particular.

Our second contribution in this paper is to show the combination of extended and simultaneous sources, since, similarly to the scenarios mentioned above, the simultaneous sources is not entirely suitable to accelerate FWI when extended sources are used. In particular, we adjust the method to the low-rank source extensions that we suggest in this paper.

The rest of this paper is organized as follows. In Section \ref{sec:backgroud} we introduce some background and notation used throughout this paper. In Section \ref{sec:LowRankExtended} we introduce our main contribution: low-rank and sparse extended sources. Then, in Section \ref{sec:ExtSimSrc} we show how to merge the extended sources method with simultaneous sources. Following that, we present numerical results in Section \ref{sec:Results}.

\section{Preliminaries and background}\label{sec:backgroud}

To model the waveforms for FWI in frequency domain, we consider the discrete acoustic Helmholtz equation as a forward problem, assuming a constant density media
\begin{eqnarray}
\label{eq:Helmholtz}
%\Delta \bfu + \omega^{2}\bfm \bfu = \delta(x-x_{s}).
%\mathcal{H}(\bfm,\omega)\bfu = \Delta_{h} \bfu + \omega^{2} (1-i\bfgamma)\odot\bfm\odot\bfu = \bfq_{s}.
\mathcal{H}(\bfm,\omega)\bfu = \Delta_{h} \bfu + \omega^{2}\bfm\odot\bfu = \bfq_{s}.
\end{eqnarray}
The symbol $\Delta_{h}$ represents the discretized Laplacian on a nodal regular grid, $\bfu = \bfu(\bfm,\omega,x_{s})$ is the discrete wavefield, $\bfm > 0$ is the model for the inverse of the squared wave velocity, $\omega$ is the frequency, and the symbol $\odot$ denotes the Hadamard product. The source, $\bfq_{s}$ is assumed to be a discretization of a delta function, $\delta(x-x_{s})$, that is located at $x_{s}$. %The vector $\bfgamma>0$ is used to impose attenuation in the wave propagation model.
The equation is discretized on a finite domain and is accompanied with absorbing boundary conditions on all sides of the domain, and possible attenuation term \cite{pratt1999}. At high wave-numbers, the Helmholtz linear system \eqref{eq:Helmholtz} is very challenging to solve numerically---it is indefinite, highly ill-conditioned, and it requires a very fine mesh to accurately capture the wave behavior. To solve FWI, we require multiple solutions of \eqref{eq:Helmholtz},
which makes the solution of the inverse problem highly expensive and cumbersome \cite{JointEikFWI17} in addition to the other difficulties mentioned earlier.

In FWI, sources are located at many locations on the top part of the grid, and the waveform that is generated by each source is recorded at locations where receivers are placed. In our formulation, each observed data sample corresponding to a source and frequency $\omega$, is given by
\begin{equation}\label{eq:data}
\bfd^{\obs}(\omega,x_{s}) = \bfP_s^{\top}\bfu(\bfm_{\true},\omega,x_{s}) + \epsilon
\end{equation}
where $\bfP_{s}$ is a sampling matrix that measures the wave field $\bfu$ that is generated by the source at $x_{s}$ at the locations of receivers. The data contain noise $\epsilon$, which we assume to be i.i.d, Gaussian, and with zero mean and (co-)variance $\Sigma$.
Given data for many sources and several frequencies,
we aim to estimate the true model, $\bfm_{\true}$, by minimizing the difference between the measured and simulated data, obtained by solving \eqref{eq:Helmholtz} across all the sources and frequencies. This may be done by solving the PDE-constrained optimization problem
\begin{eqnarray}\label{eq:Phi0}
 \min_{\substack{\bfm_{L} \le \bfm \le \bfm_{H}\\\{\bfu_{sj}\}_{j=1,s=1}^{n_f\;\;n_s} }}  \Phi_\constrained(\bfm,\{\bfu_{sj}\}) &=&
\sum_{j=1}^{n_{f}}{\sum_{s=1}^{n_s}{\left\|\bfP_s^{\top} \bfu_{sj} - \bfd^{\obs}_{sj}\right\|^2_{\Sigma_{sj}^{-1}}}} + \alpha R(\bfm)\\\nonumber
\mbox{s.t.  }
 \mathcal{H}(\bfm,\omega_j)\bfu_{sj} &=& \bfq_{s}, \quad\quad\quad\quad\quad s=1,\ldots,n_{s},\ \ j=1,\ldots,n_{f}
\end{eqnarray}
where $\bfu_{sj}$ is the waveform for source $s$ and frequency $\omega_j$, which is predicted for a given model $\bfm$, according to the forward problem \eqref{eq:Helmholtz}. The data terms $\bfd^{\obs}_{sj} = \bfd^{\obs}(\omega_j,x_{s})$ are the corresponding observed data as in \eqref{eq:data}.

We use the upper and lower bounds $\bfm_{H}>\bfm_{L}>0$, and a regularization term $R(\bfm)$, which is accompanied by a parameter $\alpha>0$.
These promote prior information in the inversion, and help us find a reasonable solution to the otherwise ill-posed problem  \cite{vogel2002computational,somersalo2004statistical}. To this end, we assume $\bfm$ to be a layered model, and choose $R$ to promote piece-wise smooth functions like the total variation regularization term \cite{rudin1992nonlinear}. We elaborate on these choices later in the results section.

Since we have many sources and several frequencies, the solution of the inverse problem at large scales requires parallel/distributed software and resources. Typically, the solution will be distributed to workers according to frequencies, sources, and even sub-domains, and some components (e.g., the gradient for $\bfm$) will be assembled on a master worker or a small group of workers. For more information, and an open-source code, see \cite{jInv17} and references therein.

\subsection{The method of frequency continuation}\label{sec:freqCont}
As stated before, reaching local minima is a major problem in solving FWI. The first step in trying to avoid local minima is adopting a method called frequency continuation, which has proven to be effective in \cite{pratt1998gauss, pratt1999}.
Frequency continuation is obtained by first approximately solving the problem \eqref{eq:Phi0} (or one of the equivalent formulations that we show later) using the lower frequencies only in its cost function, to build a smooth approximation of the velocity model.
We then add more and more frequencies and approximately solve the problems, each time starting from the previously obtained model, until we cover all of the frequencies. Algorithm \ref{alg:FreqCont} summarizes the ``window-wise'' frequency continuation approach that we use in this work, where we consider a window of frequencies at each time, instead of all of them. Furthermore, in this work we use the Gauss-Newton (GN) method to approximately minimize the corresponding $\Phi$ in every frequency continuation iteration \cite{pratt1998gauss}. We briefly present the projected GN later, and refer the reader to \cite{jInv17,JointEikFWI17} for a more detailed description of GN using similar notation. Projected Conjugate Gradients (CG) is used for the inner Newton problems.

\begin{algorithm}
 \# Assume $\omega_{1}<...<\omega_{n_{f}}$.\\
 \# $ws$: Window size of frequencies that we work on each time.\\
 \# $i_{start}$, $i_{end}$ - Initial and final frequencies to consider. $i_{start} \geq 1$, $i_{end} \leq n_f$.\\
 Initialize $\bfm$ by some reference model (or have it from a previous cycle).\\
 \For{$i=i_{start}:i_{end}$} {
   Approximately solve $\Phi$ using data for $\omega_{max\{i-ws,1\}},...,\omega_{i}$, starting from a previous model $\bfm$.
 }
 \caption{Frequency continuation}\label{alg:FreqCont}
\end{algorithm}

While having good results, the frequency continuation has a major obstacle - starting with a good initial guess $\bfm_{ref}$ is critical for an accurate estimation. Finding such a $\bfm_{ref}$ is not a trivial task, since our data does not correspond to low enough frequencies because of acquisition limitations. Some of the previously mentioned methods aim exactly at finding the initial guess, while in this work we offer a more general approach to try to escape local minima efficiently. %\sout{Our approach can be applied in addition to other approaches that are not relevant to the search-space expansion methods mentioned in this paper.}

\subsection{All-at-once methods}  One of the main approaches to solve \eqref{eq:Phi0} considers $\bfm$ and all the fields $\bfu_{sj}$ as variables, and handles the PDE constraints using the Lagrange multipliers approach, adding a Lagrange variable to each pair $s,j$ \cite{Haber2001AllAtOnce,metivier2017review}. However, in real life 3D scenarios, solving the problem is computationally demanding both in terms of memory and computations. The forward problem \eqref{eq:Helmholtz} requires a very fine mesh of hundreds of millions of grid points for each $\bfu_{sj}$. The inverse problem \eqref{eq:Phi0} typically includes several hundreds of sources $\bfq_s$ and tens of frequencies $\omega_j$. Handling the additional $\bfu_{sj}$ variables imposes severe memory issues at large scales, since these vectors are huge, and there are many of them. Manipulating them iteratively in the optimization process is quite cumbersome at such large scales.

\subsection{The reduced and penalized formulations}
Another common approach to solve \eqref{eq:Phi0} is by eliminating the PDE constraints, setting
\begin{equation}
\bfu_{sj} = \bfu_{sj}(\bfm) = \mathcal{H}(\bfm,\omega_j)^{-1}\bfq_s.
\end{equation}
This results in the reduced unconstrained (with respect to the PDEs) optimization problem
\begin{equation}\label{eq:Phi0reduced}
\displaystyle{\min_{\bfm_{L} \le \bfm \le \bfm_{H}}}\Phi_{\reduced}(\bfm) = \sum_{j=1}^{n_{f}}{\sum_{s=1}^{n_s}{\left\|\bfP_s^{\top} \mathcal{H}(\bfm,\omega_j)^{-1}\bfq_s - \bfd^{\obs}_{sj}\right\|^2_{\Sigma_{sj}^{-1}}}} + \alpha R(\bfm)\\
\end{equation}
for the model $\bfm$ only. This problem is equivalent to \eqref{eq:Phi0}, but it imposes different solution techniques. In particular, using traditional methods like GN \cite{pratt1998gauss}, it is possible to handle large instances of \eqref{eq:Phi0reduced} by occasional use of the disk \cite{JointEikFWI17}, instead of frequently manipulating the many large vectors $\bfu_{sj}$.

However, while the unconstrained \eqref{eq:Phi0reduced} is more feasible to handle than \eqref{eq:Phi0} at large scales, it is highly non-linear. This non-linearity partially stems from the elimination of the PDE constraints---the optimization process is forced to fulfill those constraints at all times. Consequently, \cite{van2016} suggested to relax these constraints. That is, to allow the fields $\bfu_{sj}$ not to satisfy the PDE \eqref{eq:Helmholtz}, which is weakly enforced by a penalty term using the $L_2$ norm. The resulting constrained problem is given by
\begin{eqnarray}\label{eq:PhiPenalized}
 \min_{\substack{\bfm_{L} \le \bfm \le \bfm_{H}\\\{\bfu_{sj}\}_{j=1,s=1}^{n_f\;\;n_s} }}  \Phi_\penalized(\bfm,\{\bfu_{sj}\}) &=&
\sum_{j=1}^{n_{f}}{\sum_{s=1}^{n_s}{\left\|\bfP_s^{\top} \bfu_{sj} - \bfd^{\obs}_{sj}\right\|^2_{\Sigma_{sj}^{-1}}}} +\\\nonumber &&\beta\sum_{j=1}^{n_{f}}{\sum_{s=1}^{n_s}{\left\|\mathcal{H}(\bfm,\omega_j)\bfu_{sj} - \bfq_{s}\right\|_2^2}} + \alpha R(\bfm).
\end{eqnarray}
This problem has penalty terms which essentially replace the PDE constraints, and $\beta>0$ is a penalty parameter that controls how accurately the constraints are fulfilled. This $L_2$ penalty method was further developed (along with suitable handling of bound-constraints and regularization) in \cite{aghamiry2019implementing,aghamiry2018improving,aghamiry2020robust} using the alternating direction method of multipliers (ADMM), which relieves the user from choosing a large $\beta$ to fulfil the constraints in high accuracy. These approaches, like the all-at-once method to solve  \eqref{eq:Phi0}, include the iterative updates of the vectors $\{\bfu_{sj}\}$, which require a lot of memory and are almost impractical in certain scenarios.

\subsection{The formulation of extended sources}\label{sec:ExtSrc}
A different approach, albeit in the same spirit as the penalty approach in the context of this work, enlarges the search space of the problem by extending the point-sources $\bfq_s$ \cite{huang2018volume}. Even though this work suggested the approach in a time-domain formulation, the equivalent formulation in frequency domain may result in the problem
\begin{eqnarray}\label{eq:PhiExtSrc}
\min_{\substack{\bfm_{L} \le \bfm \le  \bfm_{H} \\ \{\bfz_{s}\}_{s=1}^{n_s}}}
\Phi_{\extsrc}(\bfm, \{\bfz_{s}\}_{s=1}^{n_s}) &=&
\sum_{s=1}^{n_s}{\sum_{j=1}^{n_{f}}{\left\|\bfP_s^{\top}\mathcal{H}(\bfm,\omega_j)^{-1}\bfz_s  - \bfd^{\obs}_{sj}\right\|^2_{\Sigma_{sj}^{-1}}}} +\\ && \nonumber \beta\sum_{s=1}^{n_s}\|\bfz_{s}\|_{W_s}^2 +  \alpha R(\bfm).
\end{eqnarray}
Here, the extended sources $\bfz_s$ are introduced as variables, and are penalized by a weighted $\ell_2$ norm. The weight matrices $W_s$ do not penalize $\bfz_s$ at the sources locations $x_s$, hence setting the extended sources to be the original ones, i.e., $\bfz_s = \bfq_s$, does not penalize the objective. This is another form of enlarging the search space and relaxing the PDE constraints. Similarly to before, the variables
$\bfz_s$ are large and typically we have too many of them in 3D. We also note that here, we need to choose $\beta$ sufficiently large so that the extension of the sources vanishes at the end of the optimization. This method showed very promising results, but is expensive. In this work we wish to make it more applicable at large scales. We note that compared to \eqref{eq:Phi0} or \cite{van2016}, we have only $n_s$ unknown vectors $\bfz_s$ in addition to $\bfm$, instead of $n_s\cdot n_f$ vectors in the fields $\bfu_{sj}$. The downside of the extended sources approach compared to \eqref{eq:Phi0} or \cite{van2016}, is that the forward problem \eqref{eq:Helmholtz} needs to be solved repeatedly for the source variables $\bfz_s$.

\subsection{Simultaneous sources via split formulation}\label{sec:SimSrcSplit}
As mentioned before, one of the most effective ways to reduce the computational cost of solving inverse problems is by simultaneous sources, which is obtained by trace estimation \cite{haber2012effective}. That is, given a matrix $\bfA\in\mathbb{R}^{m \times n}$, we can approximately calculate its Frobenius norm by \cite{avron2011randomized}:
\begin{equation}\label{eq:traceEst}
    \|\bfA\|_{F}^{2} = trace(\bfA^\top\bfA) = \mathbb{E}_{x}\|\bfA\bfx\|_{2}^{2} \approx \frac{1}{p} \sum_{i=1}^{p}{\|\bfA \bfx_{i}\|_{2}^{2} }
\end{equation}
where each $\bfx_{i}\in\mathbb{R}^n$ is chosen from Radermacher distribution---each element is randomly chosen from the set $\{-1, 1\}$ with equal probability. We can reformulate in matrix notation: $\frac{1}{p} \sum_{i=1}^{p}{\|\bfA \bfx_{i}\|_{2}} = \frac{1}{p}\|\bfA\bfX\|_{F}^{2}$ where $\bfX$ is a $n \times p$ matrix whose column $i$ is $\bfx_{i}$.
In practice we calculate a norm of an $m \times p$ matrix instead of an $m \times n$ matrix, which can save significant computations when $p \ll n$.

To apply the simultaneous sources technique, we wish to approximate the data misfit terms in formulations like \eqref{eq:Phi0} or \eqref{eq:Phi0reduced}, by applying \eqref{eq:traceEst}. However, this is possible only when the operators $\bfP_s$ in these formulations do not depend on $s$, i.e., when all the receivers record the waveforms from all the sources.
The same goes for the error (co-)variance matrices $\Sigma_{sj}$.
Such requirements are not met in many realistic scenarios. The work of \cite{Liu_2018} handled this issue using a quite general approach: by splitting the data term. In this approach a new set of data variables $\hat\bfd_{sj}$ is introduced, and is defined on the union of the supports of $\bfP_s$ for all $s$---that is the union of the locations of all the receivers for all the sources. Next, $\hat\bfd_{sj}$ should be similar to the observed data $\bfd^{\obs}_{sj}$ at the locations of the receivers that actually record the waveform for each source $s$. To this end, \eqref{eq:Phi0reduced}, for example, is reformulated as
\begin{eqnarray}\label{eq:minDataComp}
&&\displaystyle{\min_{\substack{\bfm_{L} \le \bfm \le \bfm_{H} \\ \{\hat\bfd_{sj}\}_{j=1,s=1}^{n_f\;\;n_s}}}} \Phi_\reducedsplit(\bfm,\{\hat\bfd_{sj}\}) =  \\ \nonumber
&&\quad\quad\quad\sum_{j=1}^{n_{f}}{\sum_{s=1}^{n_s}{\left\|\bfP^{\top}\mathcal{H}(\bfm,\omega_j)^{-1}\bfq_s  - \hat\bfd_{sj}\right\|^2_{\Sigma_j^{-1}}  + \eta\|\hat\bfP_s^\top\hat\bfd_{sj}-\bfd^{\obs}_{sj}\|^2_{\Sigma_{sj}^{-1}}}}+  \alpha R(\bfm),
\end{eqnarray}
where $\bfP$ is an operator that projects a vector $\bfu$ onto the union of the  receivers' locations, and $\hat\bfP_s$ is defined to choose the subset of receivers for source $s$ out of that union (that is, such that $\bfP_s^{\top} = \hat\bfP_s^{\top}\bfP^{\top}$). $\eta>0$ is a data fitting parameter that needs to be chosen relatively high to ensure $\hat\bfd_{sj}$ will be similar to $\bfd^{\obs}_{sj}$ at the support of the receivers of each source $s$. Minimizing \eqref{eq:minDataComp} with respect to $\hat\bfd_{sj}$ is trivial, and minimizing it with respect to $\bfm$ is similar to \eqref{eq:Phi0reduced}, but can be applied by stochastic trace estimation more efficiently. Hence, techniques such as alternating minimization or variable projection \cite{aravkin2016quadratic} are highly favorable here. Adding a regularization to $\hat\bfd_{sj}$ like in \cite{Liu_2018} might improve this formulation and is also a subject of research.

To apply the simultaneous sources technique to \eqref{eq:minDataComp}, we first select the size of the subspace we want (denoted by $p$), then define a random matrix $\bfX$ of size $n_s\times p$ from Rademacher distribution. By \eqref{eq:traceEst} we obtain
\begin{eqnarray}
\label{eq:ExtSrcApprox}
\sum_s\|\bfP^\top\helm(\bfm, \omega_j)^{-1}\bfq_{s} - \hat{\bfd}_{s,j}\|_{\Sigma_j^{-1}}^{2} &=& \|\bfP^\top\helm(\bfm, \omega_j)^{-1}\bfQ - \hat{\bfD}_{j}\|_{\Sigma_j^{-1}}^{2} \\ \nonumber &\approx& \frac{1}{p}\|\bfP^\top\helm(\bfm, \omega_j)^{-1}\bfQ\bfX - \hat{\bfD}_{j}\bfX\|_{\Sigma_j^{-1}}^{2}
\end{eqnarray}
where $\bfQ$ is the matrix whose columns are all the sources $\bfq_s$, and $\hat\bfD_j$ is the matrix of all data unknowns $\hat\bfd_{sj}$. This effectively reduces the number of sources to $p$ in every iteration. Following this splitting, we henceforth present our methods assuming that $\bfP_s=\bfP$, and $\Sigma_{sj} = \Sigma_{j}$ for all the sources $s$, and ignore the extra data term in \eqref{eq:minDataComp}.

\section{Robust FWI with low-rank and sparse extended sources}\label{sec:LowRankExtended}
In this work we aim to relax the PDE-constraints without introducing full (memory consuming) variables such as $\bfu_{sj}$, aiming for a problem that is similar to the more memory-friendly \eqref{eq:Phi0reduced} rather than to \eqref{eq:Phi0} or \eqref{eq:PhiExtSrc}. To this end, we consider source extensions $\bfz_s$ as variables. In our formulation, the new sources are $\bfq_s + \bfz_s$ (or $\bfQ+\bfZ$ in matrix notation), and we aim that at the end of the reconstruction, $\bfz_s$ will vanish. However, since we have many sources, this again introduces a significant amount of extra variables, like \eqref{eq:PhiExtSrc}, that we wish to prevent. To achieve that, we couple between all the extended sources, and define all of them as a low-rank matrix. Also, we use the $\ell_1$ norm as penalty, and aim that the source extensions are sparse to further reduce the memory consumption. Because the vectors $\bfq_s$ are sparse (they are a discrete $\delta$ function), they do not require a lot of memory---we wish the same for the extensions $\bfz_s$. The $\ell_1$ norm gives a higher penalty for small non-zero numbers compared to the $\ell_2$ norm, and encourages sparse results with a high amount of zeros.

More explicitly, we use the matrix notation of \eqref{eq:ExtSrcApprox}, and define the source extension matrix as $\bfZ=\bfZ_1\bfZ_2$, where $\bfZ_1\in\mathbb{C}^{N\times n_{es}}$ and $\bfZ_2\in\mathbb{C}^{n_{es}\times n_s}$ are the matrices that form the low-rank decomposition of $\bfZ$.
$N$ is the total number of grid nodes in the domain, $n_s$ is the number of original sources and $n_{es}<n_s$ is the maximal rank of the extended sources matrix $\bfZ$ (essentially the number of extended sources). $n_{es}$ is chosen to be small enough such that $\bfZ_1$ can reasonably fit in memory, and the computations involving $\bfZ_1$ are reasonable. On the other hand, $\bfZ_1$ needs to be rich (wide) enough to allow the relaxation of the PDE-constraints. This is a balance that we need to manage. Our optimization problem (in matrix notation) becomes:
\begin{eqnarray}\label{eq:PhiLowRankExtSrc}
&&\quad \min_{\substack{\bfm_{L} \le \bfm \le  \bfm_{H} \\ \bfZ_1,\bfZ_2}}
\Phi_{\lowrankextsrc}(\bfm, \bfZ_1,\bfZ_2)  =\\\nonumber
&&\quad\quad\quad\sum_{j=1}^{n_{f}}{\left\|\bfP^{\top}\mathcal{H}(\bfm,\omega_j)^{-1}(\bfQ+\bfZ_1\bfZ_2)  - \bfD^{\obs}_{j}\right\|^2_{\Sigma_{j}^{-1}}} +   \beta_1\|\bfZ_1\|_1 + \frac{\beta_2}{2}\|\bfZ_2\|_F^2 +  \alpha R(\bfm).
\end{eqnarray}
$\beta_1,\beta_2>0$ are regularization parameters that controls how much we allow the $\bfZ$ matrices to change and reduce the objective. The penalty term for $\bfZ_2$ is the Frobenius norm, and the penalty for $\bfZ_1$ is the element-wise $\ell_1$ norm to promote sparsity in $\bfZ_1$. This stems from memory considerations to help cases where the required rank of $\bfZ$ needs to be high. This way, the PDE-constraints are relaxed, since we remove the demand to strictly satisfy the PDE constraints for the original sources $\bfQ$. The advantage is that we achieve that without increasing the required memory by a lot. The closely related extended sources approach of \cite{huang2018volume}, presented in \eqref{eq:PhiExtSrc} uses a full rank $\bfZ$ and a weighted $\ell_2$ norm instead of an $\ell_1$ norm. These differences have a significant practical implication in 3D, as the low-rank structure and sparsity of $\bfZ$ are beneficial because of computational and memory considerations.

\subsection{Iterative solution by alternating minimization}
We solve the objective \eqref{eq:PhiLowRankExtSrc} by alternating minimization (ALM), where we alternate between the approximate minimization of $\bfm$, $\bfZ_1$, and $\bfZ_2$ in turns. Minimizing for $\bfm$ is equivalent to minimizing \eqref{eq:Phi0reduced}, and can be obtained by the LBFGS \cite{rao2017seismic} or GN methods. Minimizing for $\bfZ_2$ can be obtained directly, as it is a minimization of a rather small quadratic function (details are given later). Minimizing for $\bfZ_1$ can be done efficiently using iterative re-weighted least squares (IRLS) or proximal methods (in particular, proximal CG \cite{zibulevsky2010l1,treister2012multilevel}). Alg. \ref{alg:FreqContALM} presents the ALM algorithm which is integrated into the standard frequency continuation procedure in Alg. \ref{alg:FreqCont}.

{\centering
\begin{minipage}{0.9\linewidth}
\begin{algorithm}[H]
\#Assume $\omega_{1}<...<\omega_{n_{f}}$.\\
 \# $ws$: Window size of frequencies that we work on each time.\\
 \# $i_{start}$, $i_{end}$ - Initial and final frequencies to consider. $i_{start} \geq 1$, $i_{end} \leq n_f$.\\
Initialize $\bfm^{(0)}$ by some reference model.\\
Initialize $\bfZ_{1}$ randomly (if there is no previous solution).\\
\For{$i=i_{start}:i_{end}$} {
    \For{$j=1:iter_{ALM}$} {
        \begin{enumerate}[leftmargin=15pt]
       \item Solve for $\bfZ_2$ given $\bfZ_1$.\label{step:chooseX}
       \item Apply CG iterations for $\bfZ_1$.
       \item Solve for $\bfZ_2$ given $\bfZ_1$.
       \item Apply a GN iteration for $\Phi(\bfm)$ using data for $\omega_{max\{i-ws,1\}},...,\omega_{i}$, \newline starting from a previous model $\bfm$.
       \item  Update $\beta_1,\beta_2$ as in Section \ref{sec:updateBeta}.
        \end{enumerate}
    }
 }
 \caption{Frequency continuation with alternating minimization.}\label{alg:FreqContALM}
\end{algorithm}
\end{minipage}
\par
}

\subsubsection{Solving for $\bfm$ using Gauss-Newton}
The solution of \eqref{eq:PhiLowRankExtSrc} with respect to $\bfm$ is hardly influenced by the extended sources, and is similar to the solution of \eqref{eq:Phi0reduced}. At each iteration $k$ of GN we obtain the linear approximation
\begin{equation}\label{eq:GNapprox}
\bfu_{s,j}(\bfm^{(k)}+\delta\bfm) \approx \bfu_{s,j}(\bfm^{(k)}) + \bfJ_{s,j}(\bfm^{(k)})\delta\bfm,
\end{equation}
where $\bfu_{sj}(\bfm) = \mathcal{H}(\bfm,\omega_j)^{-1}(\bfq_s+\bfz_s)$, and $\bfJ_{s,j} = \nabla_{\bfm} \bfu_{s,j}$ is the Jacobian matrix of the data for every source $s$ and frequency $\omega_j$. $\bfz_s$ is the $s$-th column of $\bfZ_1\bfZ_2$.
In real-life scales, the Jacobian matrix cannot be stored in memory \cite{hao2000}, but we can apply matrix-vector products with this matrix and its conjugate transpose.

At each step, we place \eqref{eq:GNapprox} in \eqref{eq:PhiLowRankExtSrc} and solve a quadratic minimization by an iterative solver where only matrix vector products of the Jacobians are computed. To obtain the GN step $\delta\bfm$, we first compute the gradient of \eqref{eq:PhiLowRankExtSrc}, which is given by
\begin{equation}\label{eq:gradpi}
\nabla_{\bfm}\Phi(\bfm^{(k)}) = \sum_{s,j} \bfJ_{sj}(\bfm^{(k)})^{\top}\bfP( \bfP^{\top} \bfu_{sj}(\bfm^{(k)}) - \bfd_{sj}^{\obs}) + \alpha \nabla_{\bfm} R (\bfm^{(k)}),
\end{equation}
and then we approximately solve the linear system
$\bfH \delta \bfm = -\nabla_{\bfm}\Phi(\bfm^{(k)}),$
where the Gauss-Newton Hessian is defined by
\begin{equation}\label{eq:Hessian}
\bfH =  \sum_{s,j} \bfJ_{sj}(\bfm^{(k)})^{\top} \bfP \bfP^{\top} \bfJ_{sj}(\bfm^{(k)}) + \alpha \nabla^{2} R(\bfm^{(k)}).
\end{equation}
Once the linear system is approximately solved, the model is updated, $\bfm \leftarrow \bfm + \mu \delta \bfm$ where $\mu$ is a line search parameter that is chosen such that the objective function is sufficiently decreased at each iteration (the Armijo rule).

The Jacobian matrix required in \eqref{eq:gradpi}-\eqref{eq:Hessian}, is given by
\begin{eqnarray}
\label{eq:fwisens}
\bfJ_{sj}(\bfm) =   -\omega^2 \mathcal{H}(\bfm,\omega_j)^{-1} \, \mbox{diag}(\mathcal{H}(\bfm,\omega_j)^{-1}\bfz_s).
\end{eqnarray}
Assuming that the fields $\bfu_{sj}$ are stored in memory (or the disk), the multiplication of this matrix with a vector requires one forward solution (for each pair of source and frequency). The fields can be stored in a rather low precision \cite{JointEikFWI17}.

\subsubsection{Solving for $\bfZ_2$}
The minimization of \eqref{eq:PhiLowRankExtSrc} with respect to $\bfZ_2$ can be obtained directly, as this is a small-sized quadratic minimization problem. Denote the residual $\bfR_j$ and temporary matrix $\bfT_j$ by
\begin{equation}\label{eq:residual}
\bfR_j = \bfD_j^{\obs} - \bfP^\top\mathcal{H}(\bfm,\omega_j)^{-1}\bfQ,\quad\quad \bfT_j = \bfP^\top\mathcal{H}(\bfm,\omega_j)^{-1}\in\mathbb{C}^{n_{r}\times N},
\end{equation}
where $n_r$ is the number of receivers. The solution for $\bfZ_2$ is given by solving $n_s$ small linear systems of size $n_{es}\times n_{es}$ with the same symmetric and positive definite matrix:
\begin{equation}\label{eq:solutionZ2}
\bfZ_2 = \left(\sum_j (\bfT_j\bfZ_1)^*\Sigma_j^{-1} (\bfT_j\bfZ_1) + \beta_2 \bfI\right)^{-1} \sum_j(\bfT_j\bfZ_1)^*\Sigma_j^{-1}\bfR_j.
\end{equation}
The residuals $\bfR_j$ and matrices $(\bfT_j\bfZ_1)$ are computed as part of the update for $\bfZ_1$, which are both kept fixed in this part. We note that $\bfZ_2$ is easily computed given $\bfZ_1$, hence, there is no need to keep its iterative state or initialize it.

\subsubsection{Solving for $\bfZ_1$}
When limited to $\bfZ_1$ only, problem \eqref{eq:PhiLowRankExtSrc} is a quadratic minimization with an $\ell_1$ penalty, which is also called the least absolute shrinkage and selection operator (LASSO) regression. This problem was originally suggested in \cite{santosa1986linear} for seismic inversion like here, but went on to be highly popular in other applications, mainly signal processing \cite{zibulevsky2010l1}. It is a well understood problem with a variety of available solvers, like IRLS, proximal CG \cite{treister2012multilevel} or SEquential Subspace OPtimization (SESOP) \cite{zibulevsky2010l1}. These three are the most suitable methods in our context, because they allow us to apply the relevant matrices as ``black-box'' operators, some of which include the solution of \eqref{eq:Helmholtz}.

Again, using the residual $\bfR_j$ and $\bfT_j$ in \eqref{eq:residual}, the problem for $\bfZ_1$ is given by:
\begin{eqnarray}\label{eq:problem_Z1}
\quad\quad\sum_{j=1}^{n_{f}}{\left\|\bfT_j\bfZ_1\bfZ_2  - \bfR_{j}\right\|^2_{\Sigma_{j}^{-1}}} +   \beta_1\|\bfZ_1\|_1.
\end{eqnarray}
Because of the coupling, we get a single  linear system inside the $\ell_2$ norm for all the unknowns in $\bfZ_1$. That is, this is not a block linear system with multiple right-hand-sides, like \eqref{eq:solutionZ2}. Using the $\otimes$ symbol for the Kronecker product, the matrix equation in \eqref{eq:problem_Z1} can also be vectorized as:
\begin{eqnarray}\label{eq:problem_Z1_V2}
\quad\quad\sum_{j=1}^{n_{f}}{\left\|(\bfZ_2^*\otimes\bfT_j)\vecsf(\bfZ_1) - \vecsf(\bfR_{j})\right\|^2_{\Sigma_{j}^{-1}}} +   \beta_1\|\vecsf(\bfZ_1)\|_1,
\end{eqnarray}
where $\vecsf()$ denotes the column-stacking of any matrix into a vector. In this work we solve the LASSO minimization using IRLS, where at each ALM iteration for $\bfZ_1$ in Alg. \eqref{alg:FreqContALM} we replace the $\ell_1$ norm with a weighted $\ell_2$ norm
\begin{equation}\label{eq:IRLS}
\|\vecsf(\bfZ_1)\|_1 \rightarrow \frac{1}{2}\|\vecsf(\bfZ_1)\|_\bfW^2, \mbox{ where } \bfW=\mbox{diag}\{(|\vecsf(\bfZ_1)| +\varepsilon)^{-1}\},
\end{equation}
where $|\vecsf(\bfZ_1)|$ denotes the stacked vector of absolute entries of $\bfZ_1$. This way, the gradient of the temporary IRLS objective smoothly approximates the gradient of \eqref{eq:problem_Z1_V2}. Then, the IRLS objective is approximately minimized by a few iterations of standard preconditioned CG for the normal equations, which directly minimizes the IRLS approximation of \eqref{eq:problem_Z1_V2} in each of its iterations. For efficiency, the operators for CG are computed in matrix form as follows
\begin{equation}\label{eq:operators}
\mathcal{OP}(\bfZ_1) = \bfT_j\bfZ_1\bfZ_2, \quad \mathcal{OP}^*(\bfR) = \bfT_j^*\bfR\bfZ_2^*,
\end{equation}
where the conjugate transposed operator $\mathcal{OP}^*(\bfR)$ is equivalent to
\begin{equation}
(\bfZ_2^*\otimes\bfT_j)^*\vecsf(\bfR) = (\bfZ_2\otimes\bfT_j^*)\vecsf(\bfR) = \vecsf(\bfT_j^*\bfR\bfZ_2^*)
\end{equation} in vectorized form. We apply a few such CG iterations (specifically, about 5 each time), as an approximate minimization for $\bfZ_1$ inside the most inner ALM algorithm. As a preconditioner, we use the matrix $\bfW$ in \eqref{eq:IRLS}.

The solution for $\bfZ_1$ and $\bfZ_2$ and update for $\bfm$ are repeated $iter_{ALM}$ times in Algorithm \eqref{alg:FreqContALM}. In this process, the LU factorization or preconditioner of $\mathcal{H}$ needed to multiply the matrices $\bfT_j$ with vectors, is obtained as part of the iteration for $\bfm$.

\subsection{The choice of the regularization parameters} \label{sec:updateBeta} The regularization parameters $\beta_1$ and $\beta_2$ play a significant role in the extended sources framework. They control how much we keep the problem \eqref{eq:PhiLowRankExtSrc} close to the original one \eqref{eq:Phi0reduced}. On the one hand, we wish to allow the matrix $\bfZ=\bfZ_1\bfZ_2$ to have significant enough values to influence the optimization process. On the other hand, we wish to keep this process close to the reality, where the extensions do not exist, since they are only artificial. The work \cite{huang2018volume} chose the parameter $\beta$ in \eqref{eq:PhiExtSrc} to be fixed, so initially, when the misfit is high with respect to $\bfm$ this parameter is relatively low, and as $\bfm$ improves with the iterations, $\beta$ becomes more significant with respect to $\bfZ$. As one can expect, we observed a similar behavior in our formulation \eqref{eq:PhiLowRankExtSrc} with $\beta_1$ and $\beta_2$. As a rule of thumb, the work \cite{huang2018volume} chose $\beta$ such that the misfit with the extended sources is about half of the misfit without the extended sources (if setting $\bfZ=0$). This is obviously a quite strong penalty, as the full $\bfZ$ can easily zero out the misfit in \eqref{eq:PhiLowRankExtSrc} almost for any $\bfm$, if not penalized. That is part of the motivation for this work. By setting a high-enough $\beta$, the work  \cite{huang2018volume} essentially limits $\bfZ$. Here, we limit it by using a low-rank structure that is much more favorable computationally, even though it does introduce some algorithmic complications, as discussed above. In our framework, we keep the ratio between $\beta_1$ and $\beta_2$ fixed---specifically, we choose $\beta_2=100\beta_1$, which was chosen based on trial and error. Throughout the iterations, we change $\beta_i$ together to keep the ratio between misfits (with and without extended sources) to be approximately 0.5. More explicitly, to keep the ratio in the section $[r_1,r_2]$ we apply the following rule:
\begin{eqnarray}
\centering
\mbox{If       }\quad\quad &&\frac{\mbox{misfit}(\bfZ_1\bfZ_2)} {\mbox{misfit}(0)} > r_2 \mbox{ then } \beta_1,\beta_2 \leftarrow \beta_1/\gamma,\beta_2/\gamma,\\
\mbox{Else if } &&\frac{\mbox{misfit}(\bfZ_1\bfZ_2)} {\mbox{misfit}(0)} < r_1 \mbox{ then } \beta_1,\beta_2 \leftarrow \beta_1\cdot\gamma,\beta_2\cdot\gamma. \nonumber
\end{eqnarray}
In this work we choose $r_1=0.3$, $r_1=0.5$, and $\gamma=1.5$.

\subsection{Computational costs}
The cost of the entire inversion is dominated by 1) the cost of the GN iterations for $\bfm$, and 2) the cost of the IRLS-CG iterations for $\bfZ_1$. The cost of the minimization for $\bfZ_2$ is quite negligible compared to the cost of the other two. Below we provide details regarding each of these components.

\subsubsection{The cost of GN (standard reduced-formulated FWI)} The cost of each GN iteration is entirely dominated by the costs of the forward solvers, and that is also the dominant cost in standard reduced FWI in  \eqref{eq:Phi0reduced}. At each gradient and Hessian-vector multiplication, we need to solve the forward problem \eqref{eq:Helmholtz} twice: once to compute the Jacobian in \eqref{eq:fwisens}, and once for its adjoint, where the adjoint Helmholtz equation is solved (that is assuming that the fields $\bfu_{sj}$ are stored). Given $\bfm^{(k)}$, if possible (e.g., in 2D) we can factorize the matrices $\mathcal{H}(\omega_j)$ for all frequencies, using a direct solver \cite{MUMPS2001,schenk2004solving}, and then only need to apply the forward-backward substitutions for each source. Solutions in 3D typically require using iterative solvers with effective preconditioners (e.g., \cite{erlangga2006novel, poulson2013parallel}) to be computationally efficient. The preconditioner often dominates the computational cost. The preconditioner setup, like the LU factorization, is applied only once per iteration. It is clear that the cost of GN is controlled by the number of frequencies we consider in a frequency continuation window, and more importantly, the number of sources. To summarize:
\begin{equation}\label{eq:costGN}
\mbox{cost(GN)}= \sum_{frequencies} \mbox{cost(LinSetup)}+\#iter(\mbox{GN.CG})\cdot 2\cdot n_s\cdot \mbox{cost(LinSolve)},
\end{equation}
where $\#iter(\mbox{GN.CG})$ is the number of inner CG iterations in GN.
Setting up the sources, i.e., multiply $\bfZ_1\bfZ_2$, may be costly if the rank of source extension $n_{es}$ is not low as we choose here.

In terms of memory, the footprint of the GN iterations is given by
\begin{equation}\label{eq:memGN}
\mbox{mem(GN)}= \sum_{frequencies} \mbox{mem(LinSetup)}+n_s\cdot\mathcal{O}(N),
\end{equation}
where $\mbox{mem(LinSetup)}$ is the memory requirements of the Helmholtz solver, and $N$ is the forward mesh size. The $\mathcal{O}(N)$ storage mainly reflects the storage of the fields $\bfu_{sj}$ that are needed in the sensitivity computations in Eq. \eqref{eq:fwisens}. If the fields are not stored, the multiplication of the sensitivities with vector requires twice the number of forward simulations.

\subsubsection{The cost of the minimization for $\bfZ_1$} The cost of the minimization for $\bfZ_1$ \emph{per extended source} is similar to that of GN. The matrices $\bfT_j$ that are defined in \eqref{eq:residual} and needed in \eqref{eq:operators} typically cannot be stored in memory.
Hence, for example, to apply $\mathcal{OP}$ to $\bfZ_1$ in \eqref{eq:operators}, we need to apply a forward simulation for each pair of extend source and frequency. This shows that the low-rank structure of the source extension $\bfZ_1\bfZ_2$ is crucial for the method to be computationally attractive. Note that the solution for $\bfZ_1$ does not require the solver or preconditioner setups for the Helmholtz operators, and can reuse the ones computed as part of GN. To summarize:
\begin{equation}\label{eq:costZ1ES}
\mbox{cost(}\bfZ_1\mbox{ solve)}=\sum_{frequencies}\#iter(\mbox{IRLS.CG})\cdot 2\cdot n_{es}\cdot \mbox{cost(LinSolve)},
\end{equation}
where $\#iter(\mbox{IRLS.CG})$ is the number of inner CG iterations in IRLS.
In terms of memory, the footprint of the $\bfZ_1$ minimization is given by
\begin{equation}\label{eq:memZ1ES}
\mbox{mem(}\bfZ_1\mbox{ solve)}= \sum_{frequencies} \mbox{mem(LinSetup)}+n_{es}\cdot\mathcal{O}(N),
\end{equation}
which can be expensive if $n_{es}$ is large. In particular, in \cite{huang2018volume} the number of extended sources equals to the number of sources $n_s$ as in \eqref{eq:PhiExtSrc}, and the cost is proportional to $n_s$ instead of $n_{es}$. In this analysis we neglect the memory saving that we can exploit from the sparsity of $\bfZ_1$, since the peak memory of the sparse solvers can reach a high percentage of the unknowns, and should be carefully controlled. In any case, this is only crucial if $n_{es}$ is high.

\textbf{Comparison:} As one can observe, the cost of the two dominant components of the algorithm are controlled by similar factors: number of CG iterations, and number of sources involved. If, as we expect, the total number of sources $n_s$ is significantly larger than the rank of extended sources $n_{es}$, then the additional computations and memory for the inversion following the low-rank source extension is low.

\section{FWI using both extended and simultaneous sources}\label{sec:ExtSimSrc}
To further ease the computational cost of the inversion we will effectively reduce the number of sources in the misfit at each GN iteration using the simultaneous sources technique presented in Section \ref{sec:SimSrcSplit}. Here we describe how to combine it with the low-rank extended sources objective \eqref{eq:PhiLowRankExtSrc}. We note that, as far as we know, no work describes the combination of simultaneous sources with the standard extended sources \eqref{eq:PhiExtSrc}. The task is not straightforward, since in some sense, the simultaneous sources technique compactly ``summarizes'' the many sources $\bfZ$ into a few by $\bfZ\bfX$. However, if all those many sources in $\bfZ$ are part of the inversion unknowns, then it is not clear how to update them based on their compactly estimated version, without investing the computations for all of them. A standard update for a full $\bfZ$ costs proportionally to $n_s$ forward linear solves---that is the type of computation that we wish to prevent.

Basically, we wish to combine equations \eqref{eq:ExtSrcApprox} and \eqref{eq:PhiLowRankExtSrc}. Given a random matrix $\bfX\in\mathbb{R}^{n_s\times p}$, the combination leads to the objective
\begin{eqnarray}\label{eq:PhiLowRankExtSimSrc}
&&\quad\;\;\quad\min_{\substack{\bfm_{L} \le \bfm \le  \bfm_{H} \\ \bfZ_1,\bfZ_2}}
\Phi_{\lowrankextsimsrc}(\bfm, \bfZ_1,\bfZ_2; \bfX)  =\\\nonumber&&
%\quad\quad\quad\quad
\frac{1}{p}\sum_{j=1}^{n_{f}}{\left\|\bfP^{\top}\mathcal{H}(\bfm,\omega_j)^{-1}(\bfQ+\bfZ_1\bfZ_2)\bfX  - \bfD^{\obs}_{j}\bfX\right\|^2_{\Sigma_{j}^{-1}}} %\\\nonumber
%&&\quad\quad\quad\quad
+   \beta_1\|\bfZ_1\|_1 + \frac{\beta_2}{2p}\|\bfZ_2\bfX\|_F^2 +  \alpha R(\bfm).
\end{eqnarray}
Essentially, compared to \eqref{eq:PhiLowRankExtSrc} we have that $\bfQ\bfX$, $\bfD^{\obs}_{j}\bfX$, and $\bfZ_2\bfX$ replace $\bfQ$ and $\bfD^{\obs}_{j}$, and $\bfZ_2$ respectively, which is similar to the standard simultaneous method in \eqref{eq:ExtSrcApprox}. By frequently changing $\bfX$ in \eqref{eq:PhiLowRankExtSimSrc}, we can essentially solve \eqref{eq:PhiLowRankExtSrc} at reduced cost.

\subsection{Alternating minimization}
To use the simultaneous sources approach, we choose a new dimensionality reduction matrix $\bfX$ in step \ref{step:chooseX} of Algorithm \ref{alg:FreqContALM}, and apply the ALM iteration with updates over $\bfm$, $\bfZ_1$ and $\bfZ_2$ based on \eqref{eq:PhiLowRankExtSimSrc}. We apply the following steps which are similar to the ones described before.

To solve for $\bfm$, the GN method is applied in the same way as before, only with the low-rank source-extension $(\bfQ+\bfZ_1\bfZ_2)\bfX$, which is easily computed in memory thanks to the sparsity of $\bfQ$ and the low-rank structure of $\bfZ_1\bfZ_2$. This part would be computationally similar also if we use standard FWI with extended sources in \eqref{eq:ExtSrcApprox}.

To account for $\bfZ_2$, we compute its counterpart $\hat\bfZ_2=\bfZ_2\bfX$ by direct minimization. That is, $\hat\bfZ_2$ replaces $\bfZ_2$ in \eqref{eq:PhiLowRankExtSrc}, and just like $\bfZ_2$ is state-less (directly computed and is not updated iteratively), $\hat\bfZ_2$ is also state-less. Moreover, it is computed using the same formula as \eqref{eq:solutionZ2}, only now with the reduced residual $\hat\bfR_j = \bfR_j\bfX$ instead of $\bfR_j$. Hence, we do not keep track of $\bfZ_2$ in the inversion, and compute its reduced version $\hat{\bfZ}_2$ directly every iteration, given $\bfX$.

The solution for $\bfZ_1$ remains the same as the minimization of \eqref{eq:problem_Z1}, only with $\hat\bfZ_2$ and $\hat\bfR_j$ given above instead of $\bfZ_2$ and $\bfR_j$, respectively. We choose the dimension of the simultaneous sources and extended sources to be similar ($p\approx n_{es}$) so that the minimization for $\bfZ_1$, even though only approximated, will not over-fit the data that is reduced by the multiplication in $\bfX$.

\subsection{Computational costs}

Because the column-dimension of $\bfZ_1$ is $n_{es}$, then the cost of the update for $\bfZ_1$ is similar to \eqref{eq:costZ1ES}. As in the case of simultaneous sources with standard FWI, the real saving is in the GN iterations. That is, we have $p$ sources instead of $n_s$, so compared to \eqref{eq:costGN} we have
\begin{equation}\label{eq:costGNES}
\mbox{cost(GN-SS)}= \sum_{frequencies} \mbox{cost(LinSetup)}+\#iter(\mbox{GN.CG})\cdot 2\cdot p\cdot \mbox{cost(LinSolve)},
\end{equation}
where $\#iter(\mbox{GN.CG})$ is the number of CG iterations in GN. Memory-wise, we have
\begin{equation}\label{eq:memGNES}
\mbox{mem(GN-SS)}= \sum_{frequencies} \mbox{mem(LinSetup)}+p\cdot\mathcal{O}(N).
\end{equation}
In short, if $p\approx n_{es}$ the solution cost for $\bfZ_1$ is proportional to the GN iterations for $\bfm$ using simultaneous sources. This is what we wanted to achieve here.

\section{Numerical Results}\label{sec:Results}

In this section we demonstrate our low-rank extended sources approach with and without the simultaneous sources technique and compare it to standard FWI for velocity model reconstruction. For the purpose of demonstration, we do not augment the inversions with other modalities or techniques known in the literature. In principle, such techniques can be used in addition to our approach in more complicated real-life experiments. We conduct our experiments on two 2D models, one is the SEG/EAGE salt model \cite{aminzadeh19973}, the other is the Marmousi model \cite{brougois1990marmousi}. For each model we include three experiments - the first is a reconstruction using standard FWI, which is obtained by minimizing the reduced formulation \eqref{eq:Phi0reduced}, using a few sweeps of Algorithm \ref{alg:FreqCont}. The second experiment is FWI with the low-rank extended sources formulation according to \eqref{eq:PhiLowRankExtSrc}, using Algorithm \ref{alg:FreqContALM}. 
The last experiment involves a reconstruction using both the extended and simultaneous sources techniques, according to \eqref{eq:PhiLowRankExtSimSrc}, again using Algorithm \ref{alg:FreqContALM}, but this time with a new random matrix $\bfX$ at every ALM iteration as explained in Section \ref{sec:ExtSimSrc}. 
For all the settings and models we display the resulting reconstructed model, and the misfit at each iteration of the GN (or ALM) methods during frequency continuation sweeps. That misfit is computed for all the sources and frequencies, and without the source extensions, regardless of the frequency window or method used.

In addition, we demonstrate the feasibility of our method in 3D, by performing an experiment on part of the SEG/EAGE Overthrust model \cite{aminzadeh19973}. In particular, we wish to demonstrate that a 3D FWI experiment with extended sources can be obtained using a rather standard workstation in terms of memory and computations.

All the code in these experiments was written in the Julia language \cite{Julia}, as part of the open source jInv framework \cite{jInv17}. Some critical parts of the code, like the LU solver (forward + backward substitution) and matrix-vector products for the forward modelling is written in C++ and is parallelized using shared memory OpenMP.  The experiments were computed on a workstation with Intel
Xeon Gold 5117 2GHz X 2 (14 cores per socket) with 256 GB RAM, running on Centos 7 Linux distribution. Our code is available online at \url{https://github.com/JuliaInv/jInvSeismic.jl}.

\subsection{Smoothing regularization terms used for the reconstructions}
The objective functions in the FWI formulations in Equations \eqref{eq:Phi0reduced}, \eqref{eq:PhiLowRankExtSrc}, and \eqref{eq:PhiLowRankExtSimSrc} contain a regularization term $R(\bfm)$. Based on the work \cite{JointEikFWI17} we apply two regularization functions - one is a high order regularization called spline smoothing, given by
\begin{equation}\label{eq:highOrderReg}
R_1(\bfm) = \|\Delta_{h}(\bfm - \bfm_{ref})\|^2_{2}.
\end{equation}
The goal of this regularization is to create a smooth model from high-frequency data. The reference model $\bfm_{ref}$ is set to be the initial guess for the inversion, and while using this regularization we keep $\bfm_{ref}$ fixed. We use \eqref{eq:highOrderReg} to obtain a good smooth model, so that the rest of the process will result in a plausible reconstruction.

The second regularization function is a standard diffusion regularization
\begin{equation}\label{eq:standardReg}
R_2(\bfm) = \|\nabla_{h}(\bfm - \bfm_{ref})\|^2_{2},
\end{equation}
where $\nabla_{h}$ represents a discretized gradient on a nodal grid.
When using this regularization we change $\bfm_{ref}$ at each GN/ALM iteration to be the resulting model. Updating $\bfm_{ref}$ encourages the model to change more at each iteration, resulting in faster convergence.
In this problem, \eqref{eq:standardReg} results in a sharp reconstruction, and is suitable to use after an initial smooth model is constructed. Total Variation regularizer is another approach to promote  sharper piece-wise constant/smooth reconstruction.

\subsection{Inversion algorithm and general settings}
In this section we describe the general setup of our inversion algorithms which is common for all experiments, and more specific details will be provided for each experiment separately.
For the inversions we use the frequency continuation strategies in Algorithms \ref{alg:FreqCont} or \ref{alg:FreqContALM}---we use 3-4 sweeps of the corresponding version depending on each setting and model, with a window size of 4 frequencies for all the sweeps.
At first, we apply frequency continuation sweeps to create an initial smooth guess, using the smoothing high order regularization \eqref{eq:highOrderReg}. The remaining sweeps are responsible for sharpening the model towards the true one, hence we use the standard regularization \eqref{eq:standardReg}. The continuation sweeps with the smoothing regularization start from the first frequency, and end at the fourth frequency ($i_{start}=1, i_{end}=4$). The next sweeps using \eqref{eq:standardReg} start from the fourth frequency, up to the last one. The number of GN or ALM iterations differ for each inversion and are noted for each experiment separately.
In each GN iteration we apply 5 projected and preconditioned CG iterations to approximately solve the inner Newton problem. As preconditioner for these iterations, we use the inverse of the Hessian of the smoothing regularization terms. This way, the low number of CG iterations also play a role in regularization, the inversion is not so sensitive to the choice of $\alpha$ (see \cite{haber2014computational} for more details on this technique). Once a direction is found, we apply a standard Armijo linesearch. To solve for $\bfZ_1$ in either \eqref{eq:PhiLowRankExtSrc} or \eqref{eq:PhiLowRankExtSimSrc}, we also apply 5 (quadratic) CG iterations, at each update for $\bfZ_1$ in an ALM iteration. Throughout all the relevant iterations, the sparsity level of $\bfZ_1$ (percentage of non-zero values) was in the range of $2\%-20\%$, contributing for the low memory footprint.

\begin{figure}
\begin{center}
 \centering
    \subfloat[True velocity model.]{{\includegraphics[width=0.5\linewidth]{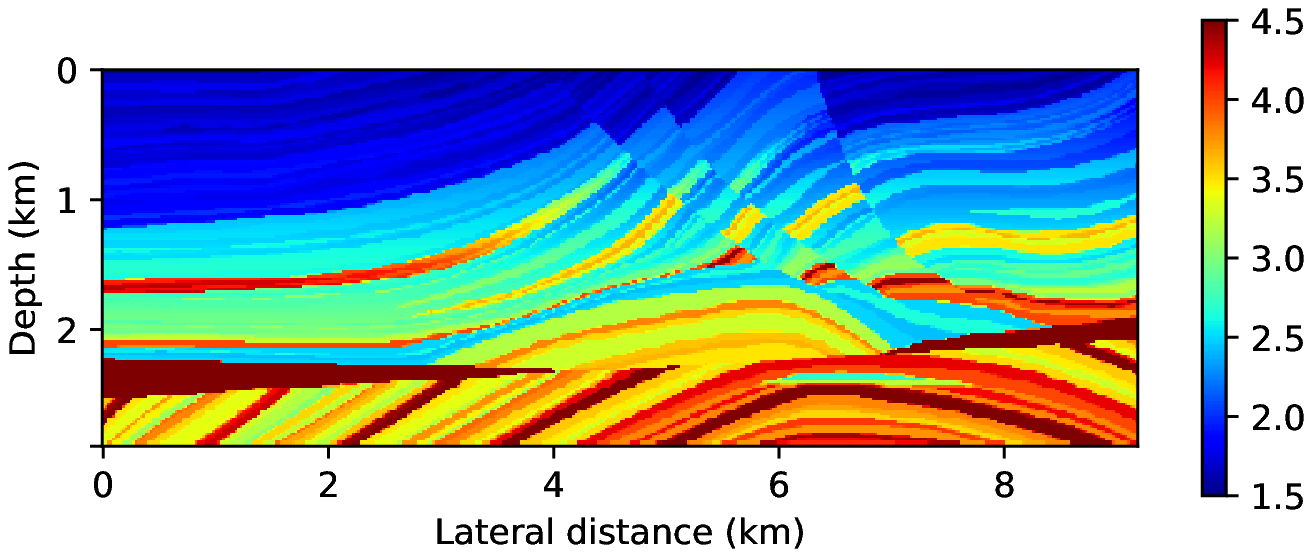}}\label{fig:marmousi_orig}}%
    \subfloat[The reference initial model.]{{\includegraphics[width=0.5\linewidth]{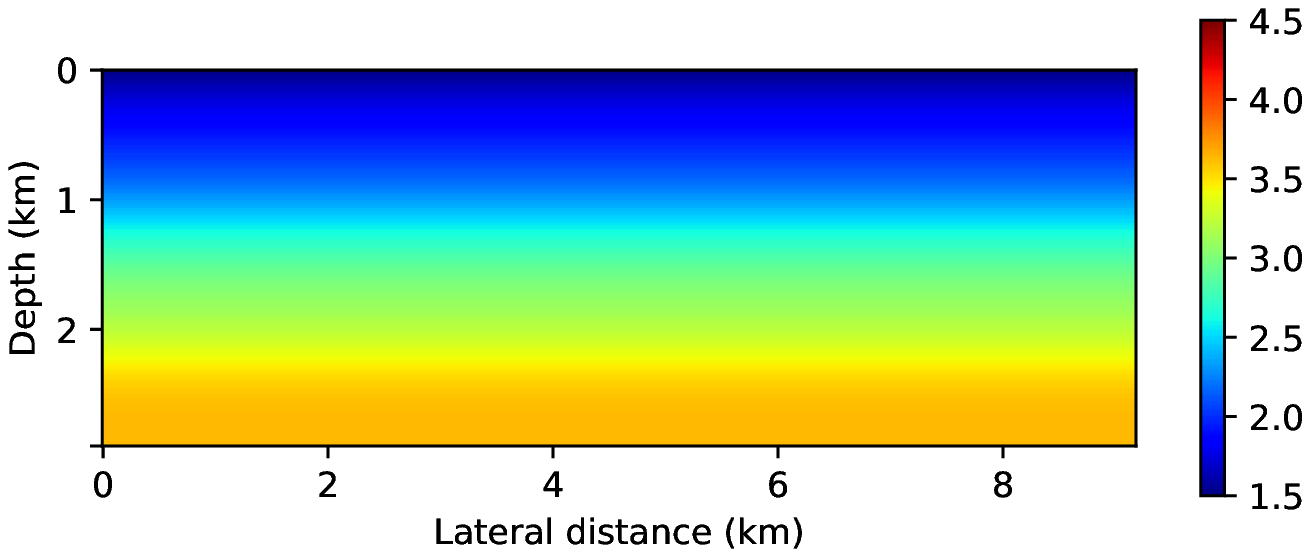}}\label{fig:marmousi_ref}}%
    \caption{The 2D Marmousi velocity model and initial guess in [km/s]. }
\end{center}
\end{figure}

\subsection{Marmousi model results}
Our first set of experiments is conducted using the Marmousi model presented in Fig. \ref{fig:marmousi_orig}. For this experiment we use a grid size of $550\times200$ representing an area of size $ 9.192 km \times 2.904 km$. We place 136 sources and 549 receivers, which are uniformly spread on the top of the grid.
The data is generated by solving \eqref{eq:Helmholtz} for frequencies $\omega_j = 2\pi f_j$, where
$$
\{f_j\} = \{3,3.5,4.0,4.5,5.0,5.5,6.5,7.5,8.5\} \; Hz,
$$
with the addition of 1\% Gaussian i.i.d noise. The initial model for all the inversions is given in Fig. \ref{fig:marmousi_ref}, and is initially used as $\bfm_{ref}$ in the smoothing regularization.

For all the Marmousi experiments we run three sweeps of frequency continuation - one sweep over the first 4 frequencies using the smoothing regularization \eqref{eq:highOrderReg}, and two additional sweeps over all the frequencies using the regularization \eqref{eq:standardReg}. All sweeps are obtained with 10 GN or ALM iterations for each outer frequency continuation iteration. This results in total of 140 GN iterations.
Then, we applied up to 100 additional GN iterations involving the four highest frequencies only, without source extensions. The additional iterations are needed to sharpen the reconstructed model, especially for the extended sources versions.

\begin{figure}
\begin{center}
 \centering
    \subfloat[Standard FWI.]{{\includegraphics[width=0.5\linewidth]{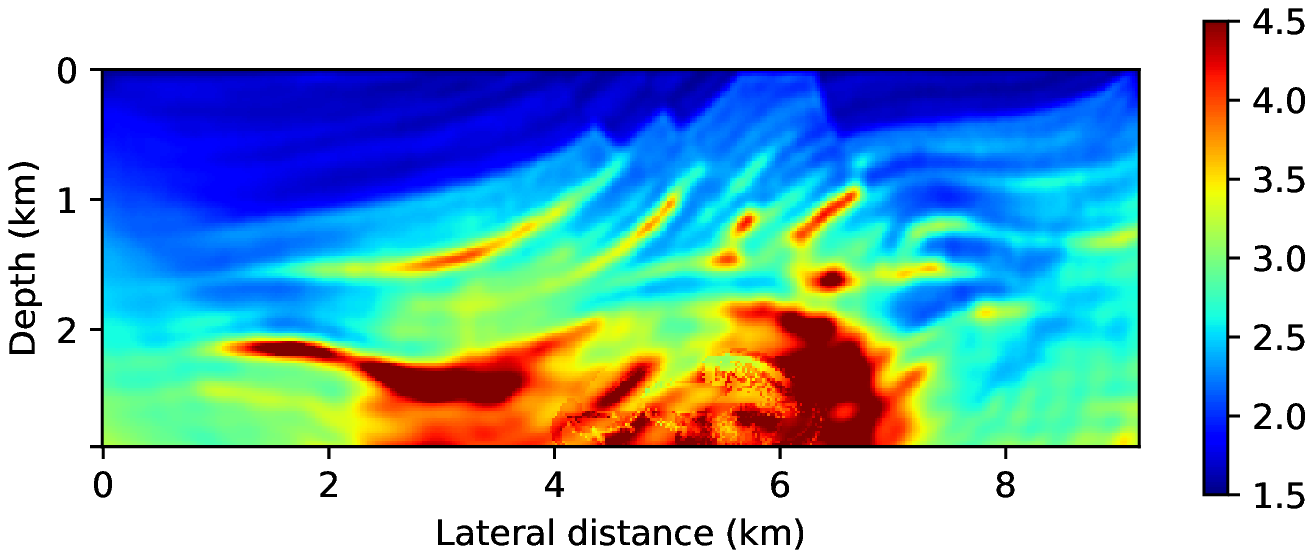}}\label{fig:marmousi_reg_res}}%
\end{center}
\begin{center}
 \centering
    \subfloat[FWI with extended sources. ]{{\includegraphics[width=0.5\linewidth]{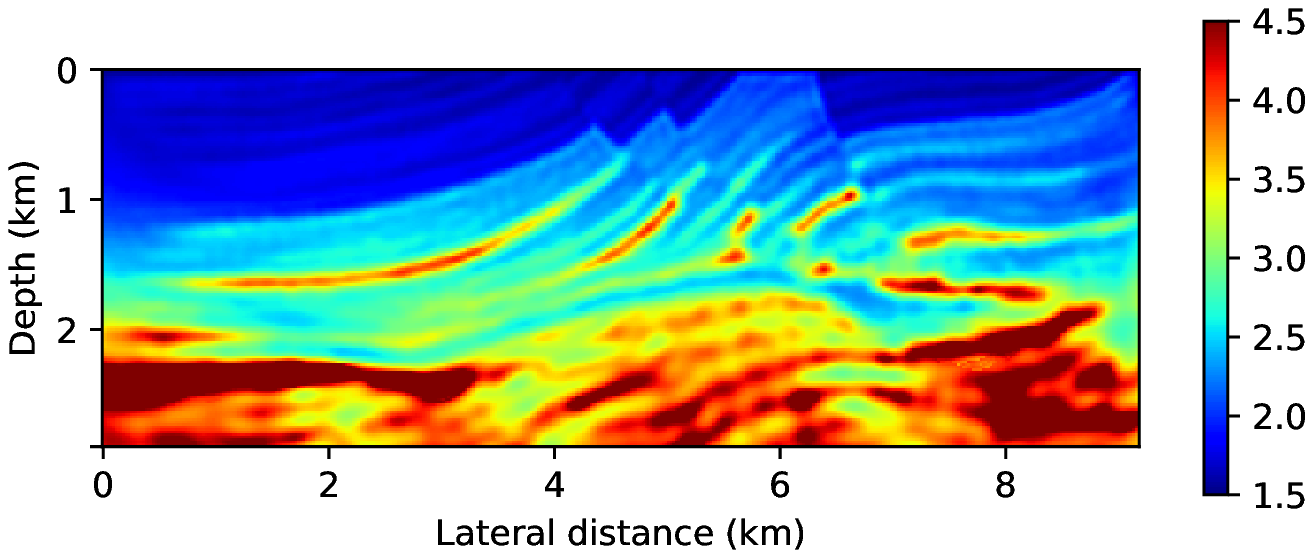}}\label{fig:marmousi_es_no_ss_res}}%
    \subfloat[FWI with extended and simultaneous sources.]{{\includegraphics[width=0.5\linewidth]{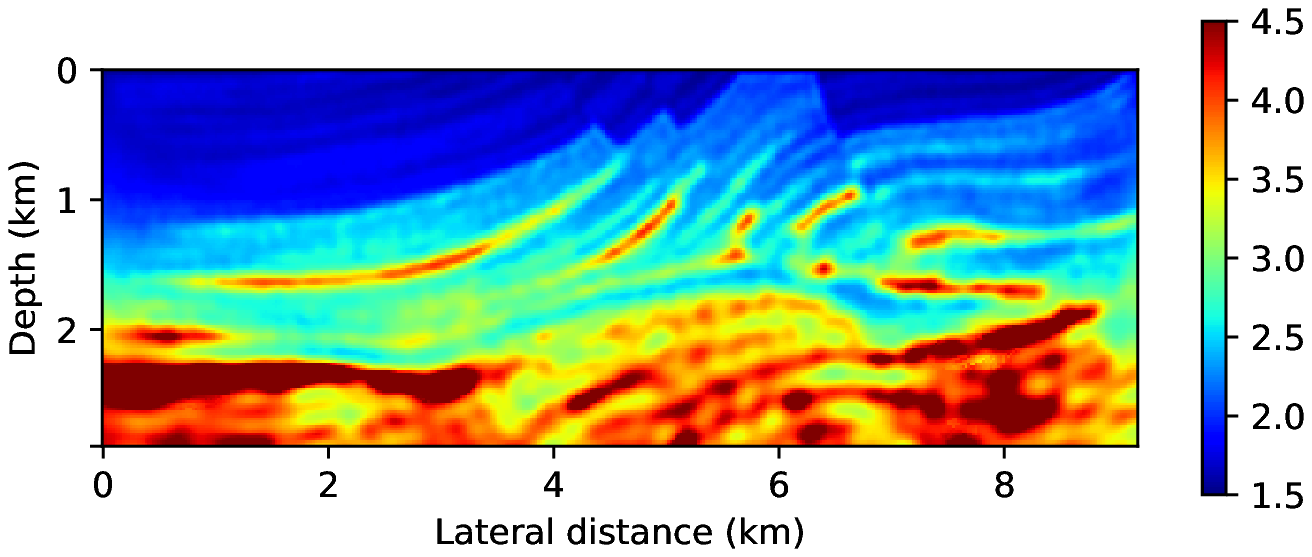}}\label{fig:marmousi_es_ss_res}}%
    \caption{Marmousi model reconstruction using the different FWI formulations.}%
\end{center}
\end{figure}

\subsubsection{Standard FWI}
The reconstructed model for the experiment is shown at Fig. \ref{fig:marmousi_reg_res}, and the misfit history plot is shown at Fig. \ref{fig:marmousi_reg_mis}.
This result converged to a local minima where the misfit value \eqref{eq:PhiPenalized} equals
6,957, and did not reconstruct the model properly. We note that in all the convergence plots, the misfit values are computed after the inversion is over for all the sources and frequencies (independently of the frequency continuation schedule), and without the source extensions.

\subsubsection{FWI with extended sources}\label{sec:Marm-FWI-ES}
For the next experiment we added the extended sources (section \ref{sec:LowRankExtended}) technique to the first frequency continuation sweep with ALM (Alg. \ref{alg:FreqContALM}). The extended sources has been used only in the first sweep since it helps with obtaining a good initial smooth guess, and once we have such a guess the standard FWI achieves good results. We set the rank of the source extensions to be $n_{es}=16$ (that is the number of columns in $\bfZ_1$). %Here we used  10 ALM iterations for the first frequency continuation sweep with high order regularization, then two more sweeps with 10 iteration and standard regularization, resulting in total of about 100 ALM (and GN) iterations.
In this experiment we chose the initial $\beta$ parameters to be $\beta_{1}=0.1, \beta_{2}=10$.
Fig. \ref{fig:marmousi_es_no_ss_res} shows the reconstructed model and Fig. \ref{fig:marmousi_es_no_ss_mis} shows the misfit history of the inversion. The experiment results in good reconstruction of the model, which is closer to the real model that the one obtained with standard FWI. The final misfit value is 1,961, which is more than half the value of the standard FWI.

\subsubsection{FWI with extended and simultaneous sources}\label{sec:MarmousiEsSs}
\begin{figure}
\begin{center}
    \subfloat[Standard FWI.]{{\includegraphics[width=0.33\linewidth]{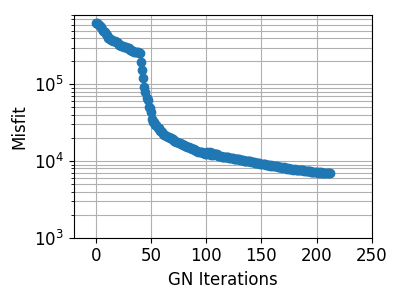}}\label{fig:marmousi_reg_mis}}%
    \subfloat[FWI with extended sources. ]{{\includegraphics[width=0.33\linewidth]{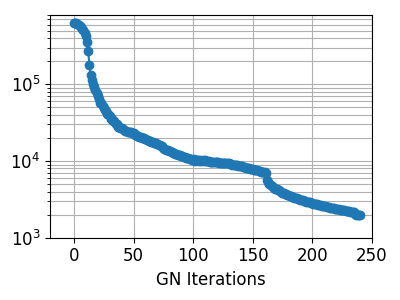}}\label{fig:marmousi_es_no_ss_mis}}%
    \subfloat[FWI with extended and simultaneous sources.]{{\includegraphics[width=0.33\linewidth]{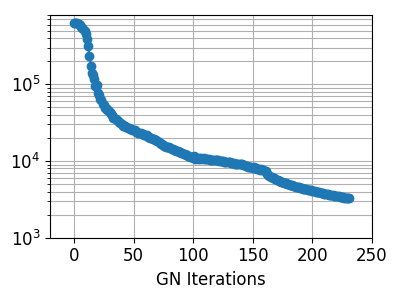}}\label{fig:marmousi_es_ss_mis}}%
    \caption{Misfit values history, using all frequencies original sources only, for the Marmousi model reconstruction.}%
\end{center}
\end{figure}

To complete the experiments for the Marmousi model, we applied the joint extended and simultaneous sources approach (Section \ref{sec:ExtSimSrc}), using the same settings as in Sec. \ref{sec:Marm-FWI-ES}.
We chose the random matrix $\bfX$ in \eqref{eq:ExtSrcApprox} to be of size $n_{s} \times p$, where $p=16$, so that in all the GN/ALM iterations, the number of sources is 16. We chose this value through trial and error, aiming at keeping $p$ small. The reconstructed model shown in Fig. \ref{fig:marmousi_es_ss_res}, and is very similar to the result in Fig. \ref{fig:marmousi_es_no_ss_res}. This similarity is also evident in the misfit history in Fig. \ref{fig:marmousi_es_ss_mis}. This shows the addition of simultaneous sources did not damage the effectiveness of the extended sources alone in converging to a better minimum.%, and only reduced the amount of computations and memory footprint to achieve it.

\subsubsection{Computational costs}
To compare the computational costs of the experiment we count the amount of forward simulations \eqref{eq:Helmholtz} that we solve in the first frequency continuation sweep of each experiment. Since the forward simulations are the most computationally expensive part of the algorithm, this gives a good comparison between the different algorithms.

\begin{center}
\begin{table}[]
    \centering
    \begin{tabular}{|c|c|c|c|}
    \hline
    Standard FWI & FWI + SS & FWI + ES & FWI + ES + SS\\
    \hline
    164,560 & 19,360 & 230,296 & 48,416 \\
    \hline
    \end{tabular}
    \caption{Comparison of computational costs for the Marmousi experiments.}
    \label{tab:marmousi_comp}
\end{table}
\end{center}

The results in Table \ref{tab:marmousi_comp} demonstrate that even with the low-rank structure, the addition of the extended sources (denoted as FWI+ES) increases the computational costs quite significantly compared to standard FWI, and especially compared to FWI with simultaneous sources (denoted FWI+SS). That is partially because of the residual computation ($\bfR_j$ in Eq. \eqref{eq:residual}), which is obtained for the full set of sources $\bfQ$. However, with the addition of the simultaneous sources (denoted as FWI+ES+SS) the cost is reduced significantly. The most computationally effective algorithm is FWI with simultaneous sources only (denoted FWI+SS), which we do not demonstrate here because its result is similar to standard FWI. Here we demonstrate that our FWI+ES+SS version only roughly doubles the cost of FWI+SS, which, just like FWI, does not involve an expanded search space and therefore is less robust.

\subsection{SEG/EAGE salt model results}
\begin{figure}
\begin{center}
 \centering
    \subfloat[True SEG/EAGE salt model.]{{\includegraphics[width=0.48\linewidth]{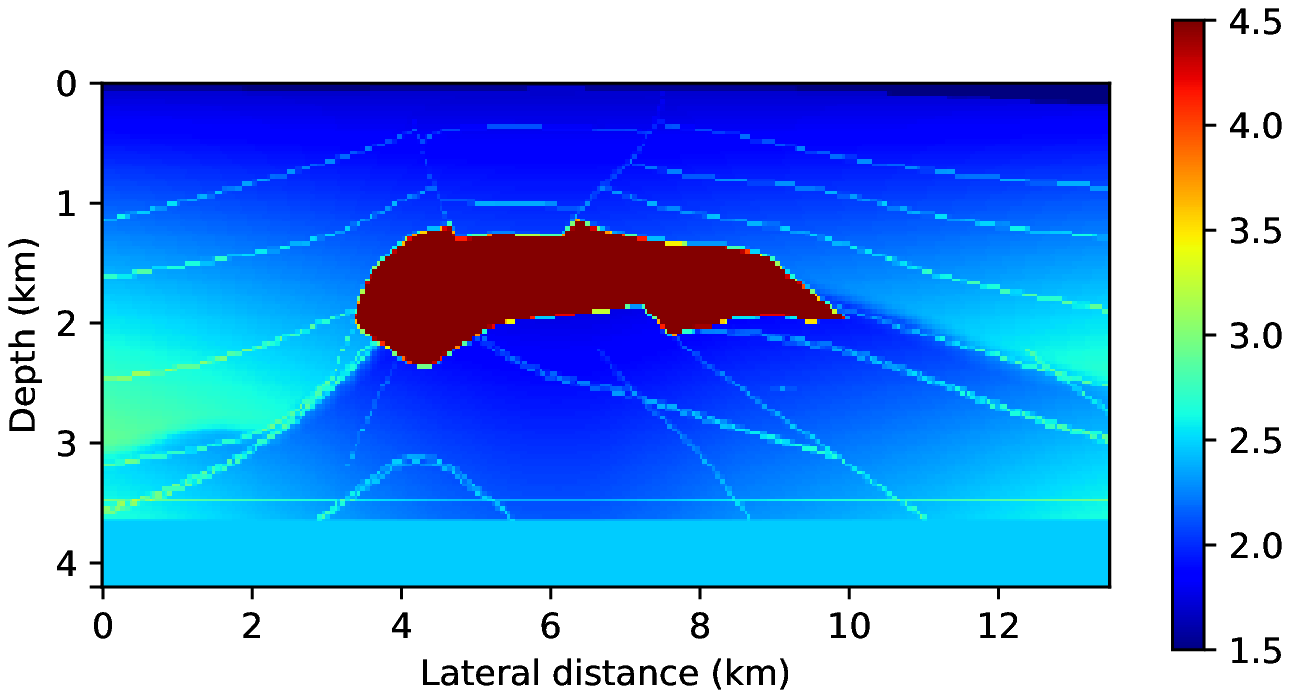}}\label{fig:seg_true}}
    \subfloat[The initial reference model.]{{\includegraphics[width=0.48\linewidth]{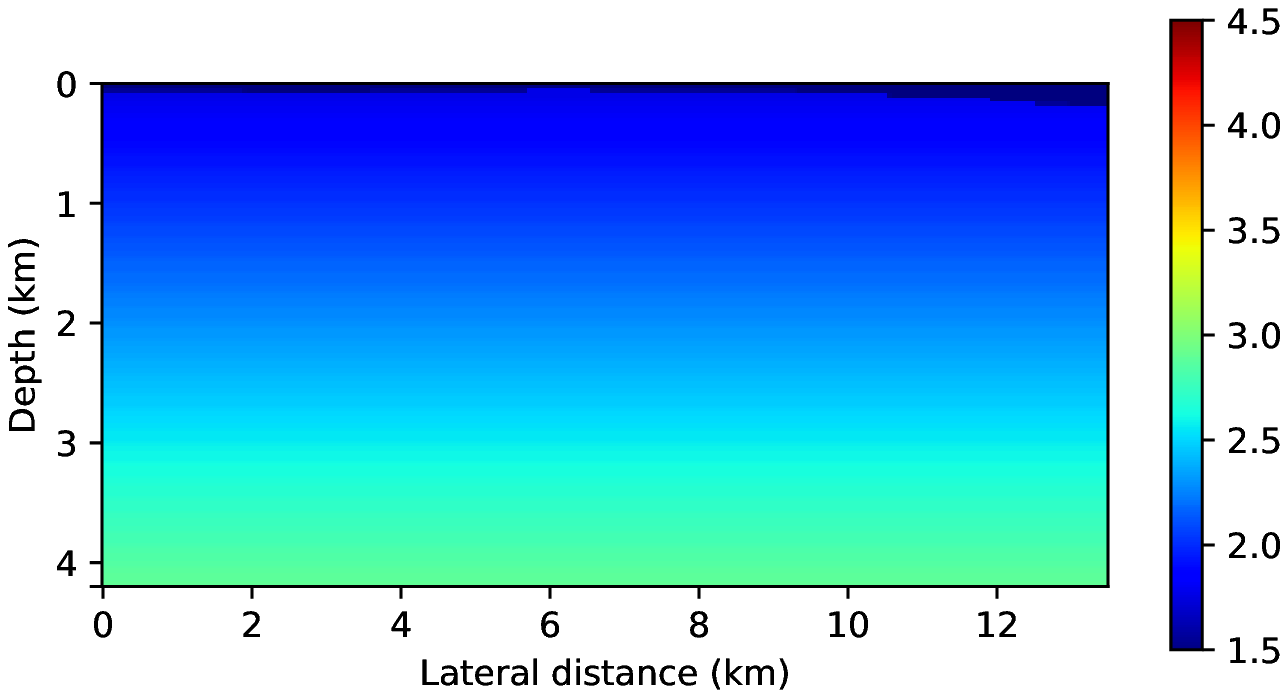}}\label{fig:seg_ref}}%
    \caption{The 2D SEG/EAGE salt velocity model and initial guess in [km/s].}
\end{center}
\end{figure}

Our second batch of experiments involves the SEG/EAGE salt model, presented in Fig. \ref{fig:seg_true}. The model is described by a $ 600 \times  300 $  grid, representing an area of size $ 13.5 km \times 4.2 km $. We placed 119 sources and 599 receivers uniformly spread at the top of the domain grid.
The data are generated by first solving \eqref{eq:Helmholtz} for frequencies $\omega_j = 2\pi f_j$, where
$$\{f_j\} = \{3,3.3,3.6,3.9,4.2,4.5,5,5.5,6.5\} \; Hz.$$
We then add 1\% Gaussian noise to the data. The initial model for all the inversions is given in Fig. \ref{fig:seg_ref}, and is initially used as $\bfm_{ref}$ in the smoothing regularization.

For the SEG/EAGE salt model we apply four frequency continuation sweeps - two using the first 4 frequencies with the regularization \eqref{eq:highOrderReg}, and two using all frequencies with the regularization \eqref{eq:standardReg} to obtain a sharp model. For the first two sweeps we used 20 GN iterations per outer iteration (with 7 CG iterations in each inner iteration), and for the last two we used 15 GN iterations (with 5 CG iterations in each, as in the rest of the configurations).
To finalize the inversion, we applied up to 100 additional GN iterations involving the four highest frequencies, without source extensions where relevant. These are used to sharpen the reconstructed model, especially for the extended sources versions.

\begin{figure}
\begin{center}
    \subfloat[Standard FWI.]{{\includegraphics[width=0.48\linewidth]{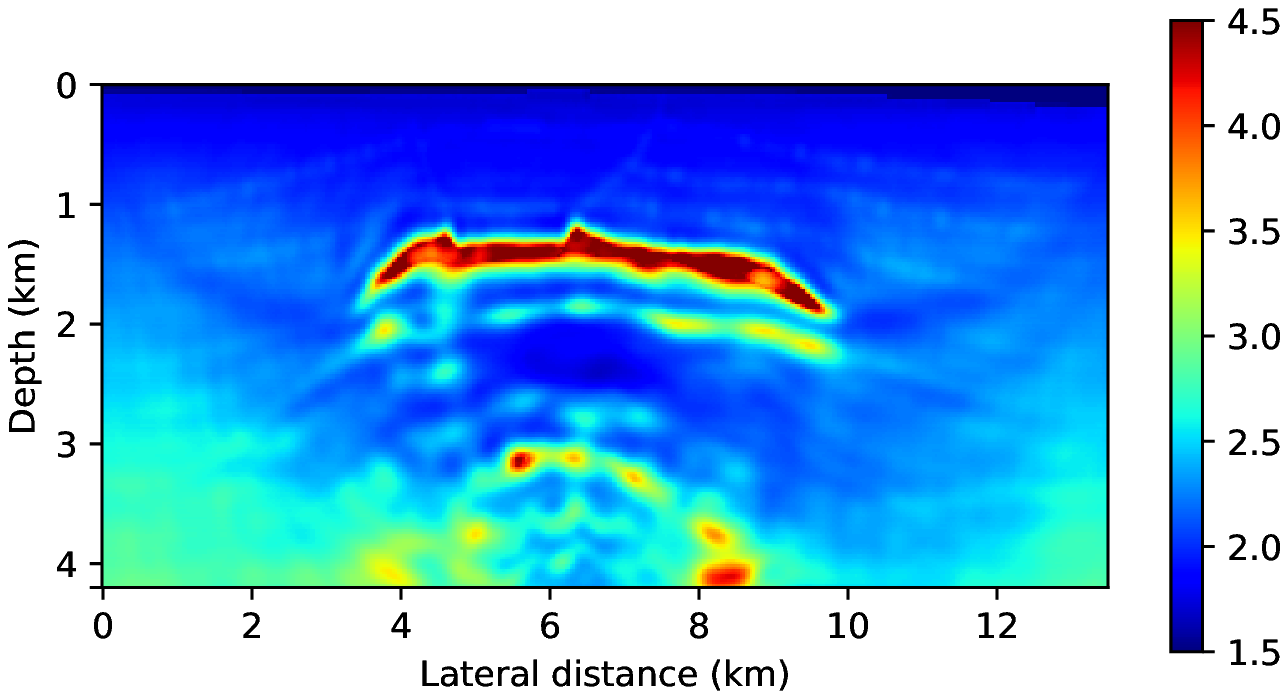}}\label{fig:seg_reg_res}}%
\end{center}
\begin{center}
    \subfloat[FWI with extended sources. ]{{\includegraphics[width=0.48\linewidth]{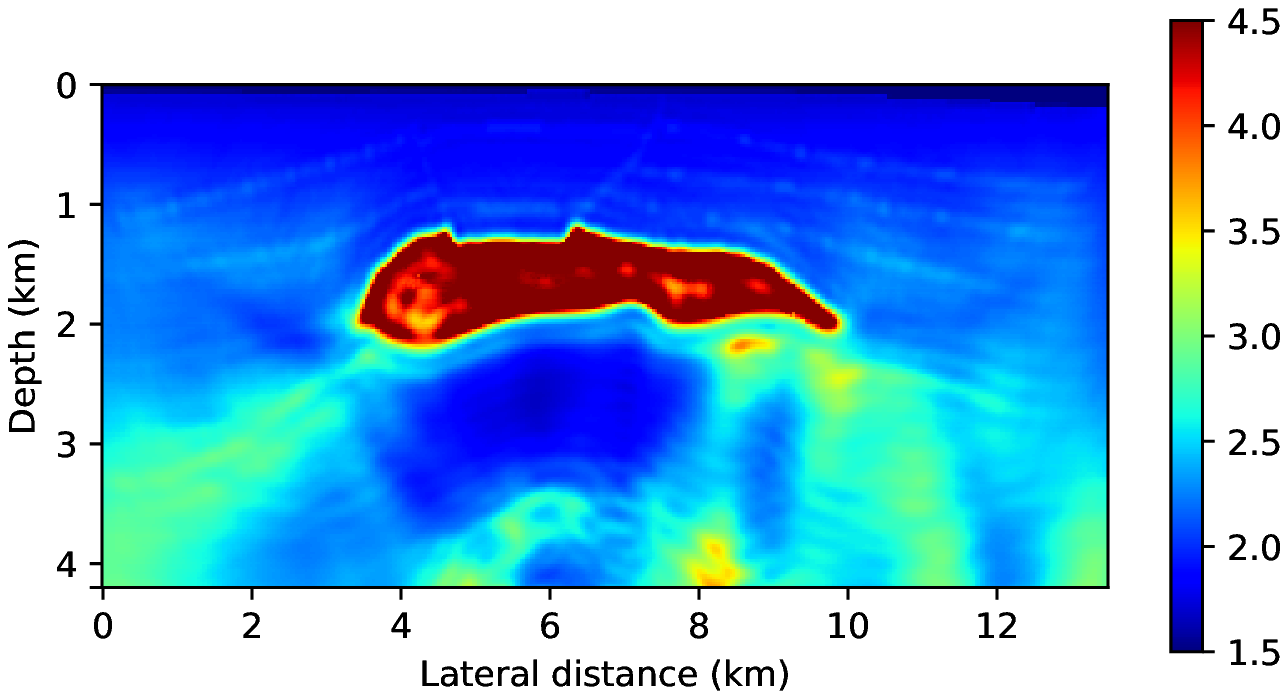}}\label{fig:seg_es_no_ss_res}}%
    \subfloat[FWI with extended and simultaneous sources.]{{\includegraphics[width=0.48\linewidth]{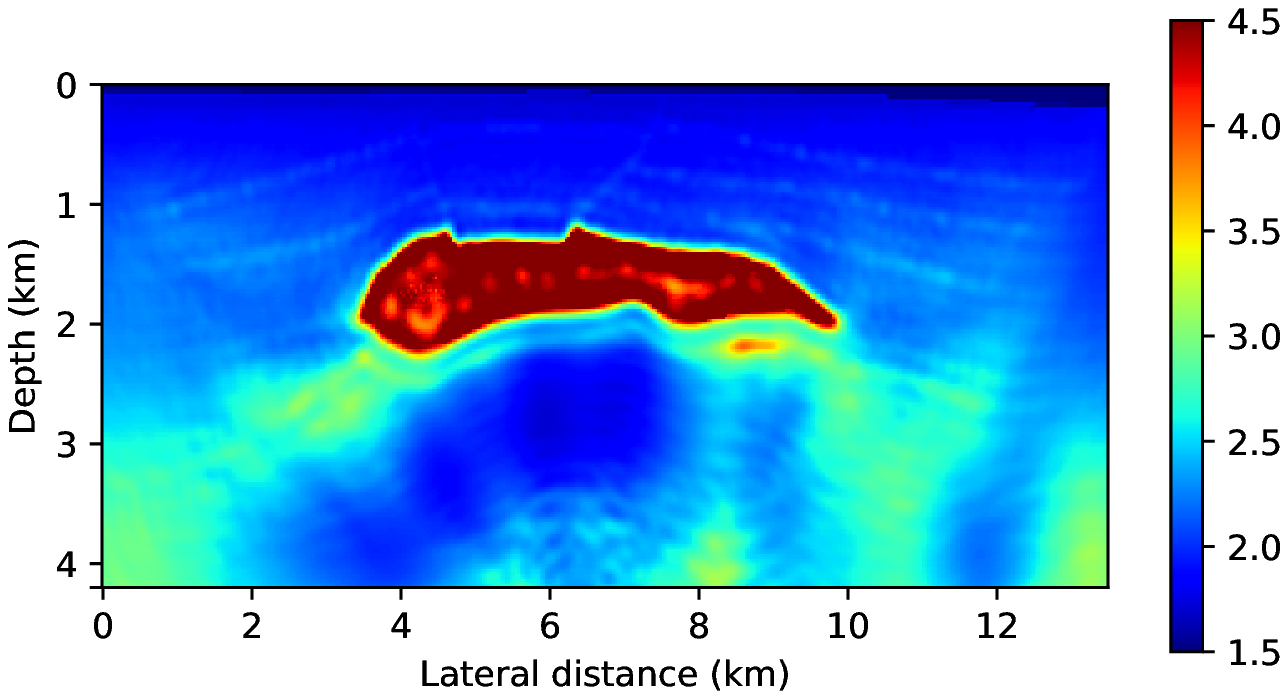}}\label{fig:seg_es_ss_res}}%
    \caption{SEG/EAGE salt model reconstruction using the different FWI formulations.}%
\end{center}
\end{figure}

\subsubsection{Standard FWI}
The reconstructed model for the experiment is shown in Fig. \ref{fig:seg_reg_res}, and the misfit history plot is shown in Fig. \ref{fig:seg_reg_mis}.
The FWI seems to only reconstruct the upper part of the salt block, and is missing the lower part. This is a typical behavior of FWI that we wish to overcome. The final misfit value here was 2278, and the iterations stagnated.

\subsubsection{FWI with extended sources}
\begin{figure}
\begin{center}
    \subfloat[Standard FWI.]{{\includegraphics[width=0.33\linewidth]{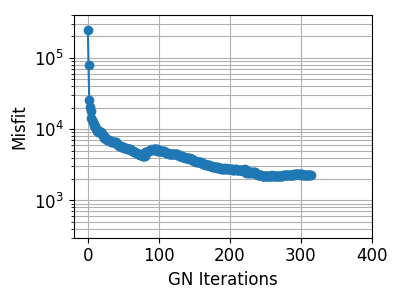}}\label{fig:seg_reg_mis}}%
    \subfloat[FWI with  extended sources. ]{{\includegraphics[width=0.33\linewidth]{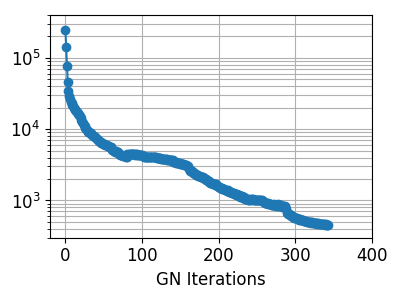}}\label{fig:seg_es_no_ss_mis}}%
    \subfloat[FWI with extended and simultaneous sources.]{{\includegraphics[width=0.33\linewidth]{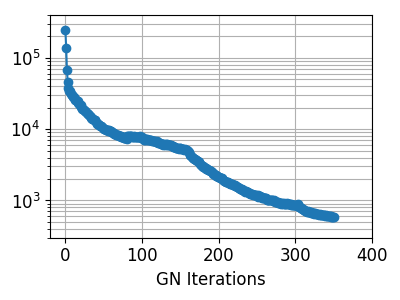}}\label{fig:seg_es_ss_mis}}%
    \caption{Misfit values history, using all frequencies and original sources, for the SEG/EAGE salt model reconstruction.}%
\end{center}
\end{figure}

In this experiment we applied our inversion strategy with the extended sources approach, which had been applied in the first continuation sweep only. In this experiment we chose the initial $\beta$ parameters to be $\beta_{1}=0.01, \beta_{2}=1$.
The reconstructed model is shown at Fig. \ref{fig:seg_es_no_ss_res}, and Fig. \ref{fig:seg_es_no_ss_mis} shows the misfit history of the inversion. The addition of the extended sources resulted in recovering the whole salt block, while keeping the extended sources at low rank. The misfit value at the last iteration was 453 - much lower than with standard FWI.

\subsubsection{FWI with extended sources and simultaneous sources}
In the last experiment for the SEG/EAGE salt model, we use both extended and simultaneous sources, with the same parameters as in the previous section.
We chose the $\bfX$ in \eqref{eq:ExtSrcApprox} to be of size $n_{s} \times p$, with $p=16$. As before, we see that the reconstructed model, shown in Fig. \ref{fig:seg_es_ss_res}, is very similar to the result in Fig. \ref{fig:seg_es_no_ss_res}. Furthermore, the misfit plots for the current run (\ref{fig:seg_es_ss_mis}) and previous run (\ref{fig:seg_es_no_ss_mis}) are similar as well, with final misfits of 591 versus 453, respectively. Like in the Marmousi experiment, the addition of the simultaneous sources hardly affected the final reconstruction, and improved the computational efficiency.

\subsubsection{Computational costs}
To compare the computational costs involved in the SEG/EAGE salt model experiments for the different algorithms, we again count the amount of forward simulations \eqref{eq:Helmholtz} we applied in the first frequency continuation sweep. The results in Table \ref{tab:seg_comp} show the same trend as in the Marmousi case. The extended sources increases the computational costs, while the addition of the simultaneous sources reduces the cost drastically.

\begin{center}
\begin{table}[]
    \centering
    \begin{tabular}{|c|c|c|c|}
    \hline
    Standard FWI & FWI + SS & FWI + ES & FWI + ES + SS\\
    \hline
    285,600 & 38,400 & 418,102 & 108,640 \\
    \hline
    \end{tabular}
    \caption{Computational costs comparison for the SEG/EAGE salt model experiments.}
    \label{tab:seg_comp}
\end{table}
\end{center}

\subsection{SEG/EAGE Overthrust 3D model results}
\begin{figure}
\begin{center}
 \centering
    \subfloat[True velocity model.]{{\includegraphics[trim={0 0 0 30.0},clip,width=0.33\linewidth]{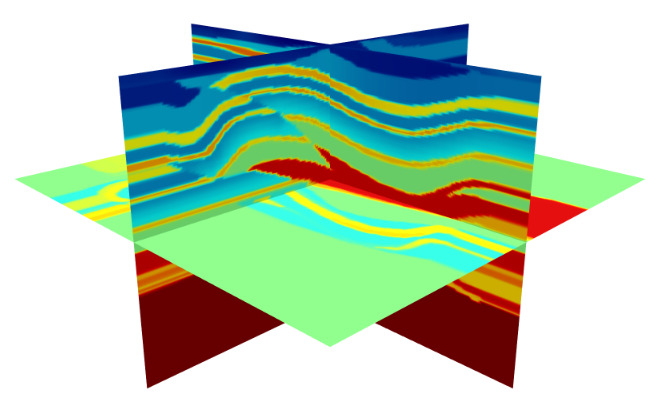}}\label{fig:overthrust_orig}}%
    \subfloat[The reference initial model.]{{\includegraphics[trim={0 0 0 30.0},clip,width=0.33\linewidth]{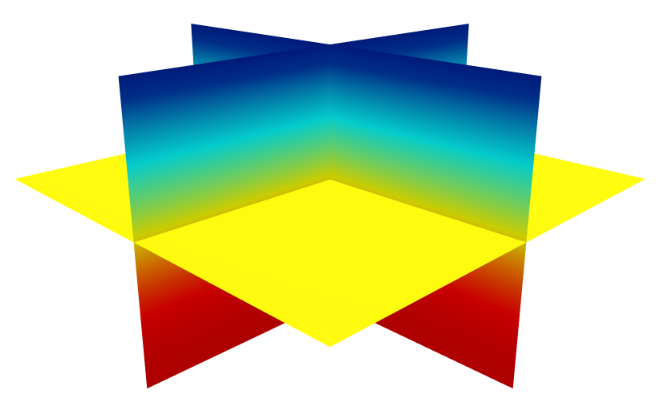}}\label{fig:overthrust_ref}}%
    \subfloat[Overthrust model reconstruction using FWI with extended and simultaneous sources.] {{\includegraphics[trim={0 0 0 30.0},clip,width=0.33\linewidth]{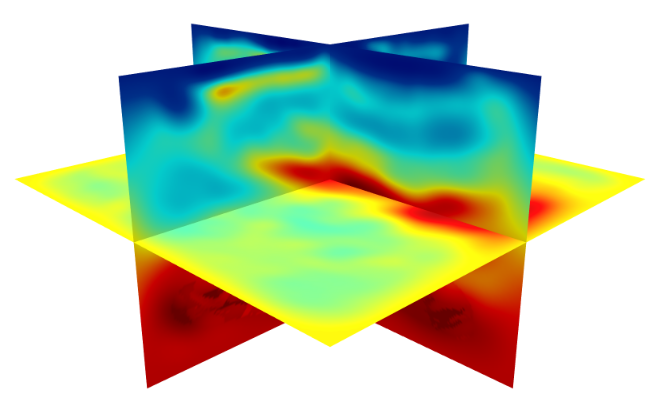}}\label{fig:overthrust_es_ss}}
    \caption{The 3D Overthrust velocity model and initial guess in [km/s]. }
\end{center}
\end{figure}

Our final experiment is applied to the central part of the 3D SEG/EAGE Overthrust model, presented in Fig. \ref{fig:overthrust_orig}. The model is discretized on a grid of size $172 \times 172 \times 108$, representing an area of size $7.5km \times 7.5km \times 4.65km$ (the center of the original model). We place 289 sources and 1849 receivers which are spread uniformly in a 2D array on the top of the grid. The data is generated by solving \eqref{eq:Helmholtz} for frequencies $\omega_j = 2\pi f_j$, where:
$$\{f_j\} = \{2.5, 3.0, 3.5, 4.0, 5.0\} \; Hz,$$
and adding white Gaussian noise of std $1\%$ of the magnitude of the data. Starting from the initial reference model in Fig. \ref{fig:overthrust_ref}, we applied 3 cycles of frequency continuation using our method with extended and simultaneous sources. In this experiment we did not run the other configurations, as they required too extensive time and careful management of memory swaps (between RAM and the disk) for our resources. For the first cycle we used the smoothing regularization with the lower 4 frequencies to obtain a smooth starting model for the next cycles. The next two cycles were applied using the standard regularization, starting from the 4-th frequency. Those cycles were applied without extended sources (FWI + simultaneous sources alone), since the model obtained after the first cycle was sufficient as a smooth guess. The resulting reconstruction presented in Fig. \ref{fig:overthrust_es_ss}. We notice that the reconstructed model, while not sharp enough, did manage to catch the important structures of the true model. The reason for the rather smooth reconstruction is that the highest frequency we used is only 5Hz to keep the model size reasonable---higher frequencies would require larger grids, and a significantly more expensive inversion.

\section{Conclusion}

In this work we aimed to improve on recent approaches for solving PDE-constrained optimization problems---approaches that expand the search space in order to relieve the non-linearity of the objective.
In particular, we considered the recent extended sources approach for FWI, and suggested a new reduced version of this problem, where we couple the source extensions as a low-rank matrix. We also showed that it is possible to accelerate the minimization of our (source-extended) objective function by using simultaneous sources, which reduces both memory and calculation costs. Unlike the previous full-rank approach, ours does not require the additional computations or memory for all the sources, which overrides the advantages of simultaneous sources, and is prohibitively expensive in large scales. Therefore, our approach is more applicable in real life 3D scenarios.

Our results showed that it is possible to combine the extended sources and the simultaneous sources as we propose. On the one hand we use the source-extension to achieve better reconstructions than the standard reduced FWI, and on the other hand we are able to enjoy manageable computations and low-memory footprint. We demonstrated our approach on two 2D models and one 3D model. The latter, in particular, demonstrates the advantage of our approach---we were able to apply the source-extended approach to a 3D problem using a rather standard workstation.

Our method has two main limitations: one is the need to a-priory choose the dimensions of both the low-rank source extension matrix and the dimension of the simultaneous sources technique. Furthermore, we found that the dimension of the latter has to be at least of the same size as the dimension of the trace estimation, to prevent over-fitting and fluctuations when solving for the source extensions.

\bibliographystyle{siam}
\bibliography{Helmholtzbib}
\end{document}